\documentstyle[12pt,epsf]{article}
\newcommand{\Date}{\ifcase\month\or January\or February\or
March \or April \or May \or June \or July \or August \or September
\or October \or November \or December \fi \ \the\day, \the\year}
\newcommand{\beq}{\begin{equation}}
\newcommand{\eeq}{\end{equation}}
\newcommand{\beqa}{\begin{eqnarray}}
\newcommand{\eeqa}{\end{eqnarray}}

\newcommand{\id}{\equiv}
\newcommand{\kr}{\kern-0.1em}
\newcommand{\kkr}{\kern-0.2em}
\newcommand{\kkkr}{\kern-0.3em}
\newcommand{\qqr}{\hspace{.5em}}
\newcommand{\qqqr}{\quad}
\newcommand{\down}{{\scriptstyle\searrow}}
\font\bsf =cmssbx10 at 12pt 
\font\bsfl =cmssdc10 at 12pt 
\font\rmVIII=cmr8 
\newcommand{\df}{\stackrel{\hbox {\rmVIII def}}{=}}
\null

\begin{document}

\thispagestyle{empty}

\begin{flushleft}
\noindent PM -- 96/19\\[-0.2cm]
April 1996\\
\end{flushleft}

\vskip0.5cm
\begin{center}

{ \Large {\bf  Anomalous thresholds and edge singularities}}\\
{\Large  {\bf in }}\\
{ \Large {\bf  Electrical Impedance  Tomography}}\\[1cm]
\end{center}

\begin{flushright}
{\it Journal of Mathematical Physics}, {\bf 37}, 4388, (1996)
\end{flushright}

\begin{center}

{ \large {\sf  Sorin Ciulli and Simona Ispas}}\\
{Laboratoire de Physique--Math\'ematique, URA 768 du CNRS}\\
{ Universit\'e\ de Montpellier II, Montpellier, France}\\[0.4cm]

{\large {\sf Michael Pidcock}}\\
{Applied Analysis Research Group, School of Computing and 
Mathematical Sciences}\\
{Oxford Brookes University, Oxford, United Kingdom}

\end{center}

\vskip 0.5cm

\noindent {\bsf Abstract:} 
 Studies of models of current flow behaviour in Electrical Impedance 
Tomography (EIT) have shown that the current density distribution varies 
extremely rapidly near the edge of the electrodes used in the technique. This 
behaviour imposes severe restrictions on the numerical techniques used in image 
reconstruction algorithms. In this paper we have considered a simple two 
dimensional case and we have shown how the theory of end point/pinch  
singularities which was developed for studying the anomalous thresholds 
encountered in elementary particle physics can be used to give a complete 
description of the analytic structure of the current density near to the edge 
of the electrodes. As a byproduct of this study it was possible to give a 
complete description of the Riemann sheet manifold of the eigenfunctions of the logarithmic kernel. These methods can be readily extended to other weakly 
singular kernels.

\newpage
\setcounter{page}{1}

\noindent {\bsf I. \qqr INTRODUCTION}

There are numerous examples of practical situations
where electric current is used to probe the interior of some object of
interest. One emerging technology which specifically uses this approach
has become known as Electrical Impedance Tomography (EIT). This is a
method of medical  and industrial imaging in which electrical currents
are applied to the surface of an object and the induced surface voltage
is measured. These data are then used to produce an image
of the conductivity distribution  in the interior of the object. An
extensive literature  exists on EIT.${}^1$
 
The particular feature of EIT which is of interest to us here is
related to the observation that in practice the electric current 
can be applied only through a finite number of electrodes--- currently 
in the range 16--64 for two dimensional applications. 
The consequences of this fact and the appropriate
mathematical modelling of the electrodes have been discussed in a
number of papers.${}^{2-4}$
 
In medical applications one of the significant problems for EIT is the
existence of a thin layer of material of high, but unknown, contact
impedance lying between the current drive electrodes and the body.
Various models have been proposed to describe   this phenomenon but 
one which has strong experimental support${}^5$ is to
suppose that on the  (finite size) electrodes
the electric potential, $\Phi$, is related   to the electric current
$\sigma (\partial \Phi / \partial n)$ by 
$$ \Phi + {\cal Z}  \sigma {\partial \Phi \over \partial n} = V  $$
where $\sigma$ is the conductivity just below the electrode,
${\cal Z}$ is the contact impedance and $V$ is the potential of the 
electrode (a constant). The  induced voltage, $\Phi$, is found by 
making measurements on high impedance electrodes, also attached to 
the surface.

This model has been studied numerically using a boundary Fourier 
technique${}^2$ for the case of constant ${\cal Z}$ and an interesting 
phenomenon which was observed was the appearance of very sharp
peaks in the current density distribution at the edge of the 
electrodes. More recently the model has been studied numerically 
for non--constant ${\cal Z}$ using the weakly singular integral 
equation described in Eq. $(\ref{e6})$ below. Although the singularity 
is weak, its existence has important consequences for the numerical 
treatment of this equation. Details of this work will be given 
elsewhere${}^6$ but the point which we wish to emphasize here is 
that the distinctive sharp peak behaviour occurs for a wide range of 
conductivity distributions and contact impedance forms. 
In Fig. 1 we show  typical results for the current density
distribution in the case of eight electrodes with an input current
on the $l$th electrode of $\cos \theta_l$.

An unwelcome  consequence of the 
sharpness of these peaks is that the  direct numerical modelling of 
the potential with the finite element method has become an excessively 
substantial task due to the high number of mesh points needed near to
the edge of the electrodes in order to accommodate 
the rapid variation of the normal derivative of the potential. 
The aim of this article is to give an explicit analytic description 
of these singularities in a form  which should substantially 
improve the speed of the numerical computation.

Since the appearance of the peaks is a boundary  phenomenon
which is very little influenced by the actual values of the 
conductivity $\sigma$  inside the  disk---see the discussion 
in Section III about the {\it dominant singular integral 
equation}---we shall focus our attention on the constant 
$\sigma$ case. In this case the governing partial differential equation
is Laplace's equation and our problem becomes one in potential theory.
Consequently, we shall study the  electrode model defined above for 
the standard domain of the unit disk.  The importance of the 
unit disk stems from the fact that the potential problem for any 
simply connected two dimensional domain  can be reduced 
to the unit disk by an appropriate conformal mapping.

\begin{figure}[t]
\begin{center}
\input{fig1}
\end{center}
\begin{quote}
{FIG. 1. {\small The current density distribution obtained 
solving Eq. (\ref{e6}) numerically  for the case of 8 electrodes, 
each having a contact impedance equal to 0.22.}}
\end{quote}
\end{figure} 

In earlier investigations of  these peaks it had been shown${}^{4}$ 
that this problem can be solved explicitly
in the {\it zero} contact impedance case 
(${\cal Z}   \equiv    0$),
for the case of  two electrodes. In this specific case 
one finds that near the edge of the electrodes 
\begin{eqnarray*}
{\partial \Phi \over \partial n} \sim {1 \over \sqrt{x}} 
\end{eqnarray*}
where $x$ is the distance,  along the boundary, from  the edge of 
the electrode.  Thus the normal derivative of $\Phi$ becomes 
infinite at the edge of the electrodes. 
However if the contact impedance ${\cal Z}$ is not zero, although 
${\partial \Phi / \partial n}$ still has sharp peaks at the edge of 
the electrodes (see Fig. 1), it  remains  finite since  both
${\Phi}$ and $V$ are finite and  since  ${\partial \Phi /
\partial n}  \equiv   (V   -   \Phi )/({\cal Z}\sigma)$. 
This shows that the nature of the singularities in the 
${\cal Z}   \not=   0$ case  {\it cannot} be 
obtained from that encountered in the soluble 
${\cal Z}  =   0$ model. Thus, in order to
have a correct understanding of these singularities some 
deeper investigations are necessary and this represents the 
main goal of the present paper.

\bigskip
\noindent{\bsf Overview of the paper} 

After a short description of the mathematical model in Section
II, in Section III we formulate the boundary problem as an 
integral equation.  As shown there the kernel of this integral equation 
has a weak (logarithmic) singularity,
and this has direct consequences for the singularities
of the solution near the electrodes edges. We will describe these 
singularities in terms of an asymptotic series for the potential.

In pursuing this program we shall have to step off the real axis
into the neighbouring complex plane. This will be necessary since
we will write the solution of the 
integral equation as an infinite sum of the free term and the 
eigenfunctions of the weakly singular kernel. As one knows, 
there are many examples of series uniformly and absolutely convergent 
on the real axis which cannot be differentiated term by term
[for example $\sum_{n=1} \cos (nx)/n^2$  converges uniformly, 
while the $k$th derivatives of its terms contain factors of the form 
$n^{k-2} $ which spoil any convergence], but
in deriving asymptotic expressions we will often have to 
perform this kind of operation. 
However, in contrast to what happens on a real 
interval, in the complex plane there exists a marvellous theorem which 
states that given a sequence $\{ f_i \}$ of  functions holomorphic  
in some domain $\Omega$ which converge  uniformly,  
$ f_i  \rightarrow f$, on all compact subsets of 
$\Omega$,  then (a) $f$ is a holomorphic function in $\Omega$ and 
 (b), $f_i'$ as well as  the  higher derivatives $f_i^{(n)}$
tend  uniformly  towards $f'$ and $f^{(n)}$ on any compact 
subset of $\Omega$. In some way by walking in the complex plane 
around the singularity one has a better view of what really 
happens there.

In studying the analytic properties of the free term (Section 
 IV) and of the eigenfunctions (Section  V), we shall use  
techniques similar to those from the  'pinch and end point 
singularities theory'  which was developed some time ago by 
Eden, Landshoff, Olive and Polkinghorne${}^7$ in elementary particle 
physics. However our problem is more complex than that related 
to the Feynman graphs in two respects. First we will have to consider 
moving cuts rather than moving poles, and second,
we will no longer have integrals over some explicitly given functions, 
but integrals over the  {\it a priori} unknown eigenfunctions whose 
singularities we are trying to find.

Handling infinite series can also be dangerous because spurious 
singularities may creep in as  happens, for example,  with the 
common geometric series. The proof that this does not happen 
in the neighbourhood of the singular point
is given in the first subsection of the Section  V.
The  analytic properties of the eigenfunctions and the 
recursive procedure to compute the coefficients in the asymptotic
expansions  are given in Section V B 
[see Eq. $(\ref{e67})$].  From this expansion it follows that the 
spikes of the current density near the edge of the electrodes 
are of the form
\begin{eqnarray*}
{\partial \Phi \over \partial n} \sim \sum_{m=1}^{r} 
\sum_{k=1}^{m} c_{mk} x^m \cdot \log^k (x) + 
{\cal O} (x^{r+1-\varepsilon}) + \hbox{ regular   part,} 
\end{eqnarray*}
where $x$ is the distance  along the boundary from  the edge of 
the electrode and $c_{mk}$ some real coefficients. Since the derivative
of $x\log (x)$ is $1+ \log (x)$,  this means that although these spikes 
are {\it finite}  they have {\it infinite derivatives}.

\vspace{0.3cm}
\noindent {\bsf II. \qqr THE MATHEMATICAL MODEL FOR EIT}

The usual model used to describe the forward problem in EIT 
is obtained by considering the object as consisting of isotropic 
material with conductivity distribution $\sigma$ contained in 
an open, simply connected region $\Omega$ surrounded by a 
reasonably smooth boundary $\partial\Omega$. On the surface, 
$\partial \Omega$, a number, $L$, of electrodes are
attached and electrical current is applied.
 
In this case Maxwell's equations give:
\beq
\nabla \cdot(\sigma \nabla  \Phi ) = 0 \hskip .7cm \hbox{ in}
\qqr \Omega . 
\label{e1}
\eeq
Further, the total current driven on the $l$th electrode 
$I_l =\int_{\Gamma_l} \sigma {\partial \Phi / \partial n }$ is 
a known quantity and there is no current  outflow outside the 
region covered by the electrodes, $\Gamma =\Gamma_1 \cup
\Gamma_2 \cup .... \cup \Gamma_L $. 
If we now introduce our electrode model mentioned earlier to the case
when $\sigma$ is constant and $\Omega$ is the unit disk,
the physical problem is  equivalent to the mathematical problem of
solving the following boundary value problem:
\beq
\nabla^2 \Phi (z) = 0 \hskip .7cm  \hbox{ in} 
\qqr \Omega , \label{e2}
\eeq
\begin{eqnarray}
{\partial \Phi \over \partial n}\kern-0.5em &=& \kern-0.5em 0 
\hskip 3.2cm \hbox {on } 
\partial\Omega \setminus \Gamma  ,  \nonumber\\ 
 {\partial \Phi \over \partial n}\kern-0.5em &=& \kern-0.5em {1 \over 
{\cal Z}_l (z)} [V_l - \Phi (z)]
\qqr \qqr  \hbox {on the electrode} \  \Gamma_l \subset \partial\Omega
, \label{e3}\\
\int_{\Gamma_l}\kkkr{\partial \Phi \over \partial n}
d\theta \kern-0.5em &=& \kern-0.5em I_l ,
\hskip 2.8cm l=1,...,L , \nonumber
\end{eqnarray}
for constant induced voltage, $V_l$, and   
total current driven, $I_l$, on each electrode.
Here  ${\cal Z}_l $ represents the contact (the 'skin') impedance and
$\Phi (z =  e^{i\theta} \in  \Gamma_l)$ is the potential just 
underneath 'the skin'.

\vspace{0.3cm}
\noindent {\bsf III. \qqr THE INTEGRAL EQUATION}

If the values of the normal derivative
$$ \left . \frac{\partial \Phi }
{\partial n} \right|_{z =e^{i \theta }}$$
were known everywhere on the boundary $\partial \Omega $ of the unit 
disk, we would be considering a classical Neumann problem, which 
is readily solved by means of the formula 
\beq
\left . \Phi (z) = \int_0^{2 \pi} {\cal N} (z, e^{i \theta'};0)
{\partial \Phi \over \partial n} \right|_{z'=e^{i \theta'}} d\theta' +
\hbox {const.},\label{e4}
\eeq
where the Neumann kernel ${\cal N} (z, z';0)$ is 
$${\cal N} (z, z';0) =-{1\over \pi }\log \left| z - z' \right|  ,$$
with $z'  =  e^{i \theta'}$  on the unit circle.
 
If we integrate the  kernel with the values of the
normal derivative $ {\partial \Phi / \partial n} $ on the 
unit circle, we obtain  a function  $\Phi$ which is harmonic
throughout the unit disk, which vanishes at $z = 0$ 
and which has the prescribed normal derivative values on the boundary. 
Similar kernels ${\cal N} (z, z';z_0 ) $ producing functions vanishing 
at $z= z_0 $ rather
that at $z=0$ can be written easily,${}^8$ but 
the knowledge of the normal derivative determines $\Phi$ only up to a
constant.
 
For what follows it is interesting to continue the Neumann integral
$(\ref{e4})$ up to the boundary. Since
$$ \left| e^{i \theta} -  e^{i \theta'} \right|^2
\equiv 2[1 - \cos(\theta'-\theta) ]   ,$$
if $z$ and $z'$ are of the
form $z  =   e^{i \theta}$ and $z' = e^{i \theta'}$
the Neumann kernel on the unit circle reads:
\begin{eqnarray}
 {\cal N} ( e^{i \theta}, e^{i \theta'};0)
&=& - {1\over 2\pi } \log \{ 2[1-\cos(\theta '-\theta )]\}  ,
\nonumber \\
\setcounter{equation}{5}&=& - {1\over \pi } \log \left | 2 \sin
\left({\theta' -\theta \over 2}\right) \right |  . 
\label{e5}
\end{eqnarray} 
In our case the normal derivative is known explicitly only on 
that part of the boundary which lies  between the electrodes 
(i.e., on $ \partial \Omega \backslash \Gamma $) where
$ {\partial \Phi / \partial n} $ is identical to zero.
However, since the integral representation $(\ref{e4})$ can 
be continued up to the boundary, conditions $(\ref{e3})$
yield a linear integral equation for the boundary values
$\rho (\theta ) \equiv \Phi ( e^{i \theta}) $ of the potential: 
\begin{equation}
\rho (\theta )  =  -  {1\over \pi } \sum_{l=1}^L V_l
\int_{\Gamma_l} {d \theta'  \over {\cal Z}_l (\theta') } \log \left | 
2 \sin \left({\theta' - \theta \over 2}\right)\right |
+ {1\over \pi } \sum_{l=1}^L  \int_{\Gamma_l} { d\theta'\over
{\cal Z}_l (\theta') }\log \left | 
2 \sin \left({\theta' - \theta \over 2}\right)\right |
 \rho (\theta ')  .\label{e6}
\end{equation} 
 
Although the kernel
$$ -{1\over \pi } \log \left| 2\sin \left({{\theta' - 
\theta}\over 2}\right)\right|$$
becomes infinite each time $ \theta' $ equals $\theta $,
this logarithmic singularity is weak enough to be $L^2$ integrable. The
kernel is therefore of Hilbert--Schmidt type and so one can benefit 
from all the advantages of Fredholm integral equations of the second 
kind, namely the existence   and uniqueness of an $L^2$ solution
$\rho (\theta)$.

Equation $(\ref{e6})$  may be rewritten in  a form which exhibits the 
logarithmic singularities of the kernel. Taking
$e^{i\theta} \in \Gamma_{l_0}$  we may write
\begin{eqnarray}
\rho (\theta )  &=& - {1\over \pi } \sum_{l \not= l_0} 
\int_{\Gamma_l} \log \left | 
2 \sin \left({\theta' - \theta \over 2}\right)\right |
 { {[V_l- \rho (\theta')]} \over {\cal Z}_l 
(\theta') }d \theta'
-{1\over \pi }  \int_{\Gamma_{l_0}} 
\log \left |  \frac {\sin \displaystyle{\left(\frac{\theta'- 
\theta}{2}\right)}}
{\displaystyle
\left(\frac {\theta' - \theta}{ 2}\right)} \right |
 { {[V_{l_0}- \rho (\theta ')]}
 \over {\cal Z}_{l_0} (\theta') }d \theta' 
\nonumber \\
&-& {1\over \pi } V_{l_0} \int_{\Gamma_{l_0}} {d \theta'  \over 
{\cal Z}_{l_0} (\theta') } 
\log \left | \theta' - \theta \right | +
{1\over \pi } \int_{\Gamma_{l_0}} {d \theta'  \over 
{\cal Z}_{l_0} (\theta') } 
\log \left |\theta' - \theta \right |\rho (\theta ')  . 
\label{e7}
\end{eqnarray}
In order to study the local behaviour of the solution 
$\rho (\theta)$ near the edges of the $l_0$th electrode, we 
absorb the first two terms which are continuous into
the 'free term' of the so called {\it dominant singular integral 
equation},${}^9$ which is of the  form:
\begin{eqnarray}
f(x) = g(x) &+&\lambda\int_0^1  K(x,t)   f(t) dt,
 \hskip 1cm  0\le x \le 1, \label{e8} \\
K(x,t) & \equiv & \log\left| t-x\right| .
\label{e9}
\end{eqnarray}
Here the points $x  =  0$ and 
$x  =  1$ correspond to the edges of the electrode
under consideration. Equation $(\ref{e8})$ can be readily deduced from 
$(\ref{e7})$ by a suitable change of variables and functions in the 
${\cal Z}_l (\theta)$ constant case,
but as we shall show elsewhere${}^{10}$ the discussion for 
the general case (non-constant ${\cal Z}_l$, non-constant
$\sigma$) is fairly similar. When we follow this procedure we find
that, as well as the first two terms from Eq. $(\ref{e7})$ which 
are regular, the function $g(x)$ contains the term 
$$-{1\over \pi } V_{l_0} \int_{\Gamma_{l_0}} {d \theta'  \over 
{\cal Z}_{l_0} (\theta') } 
\log \left | \theta' - \theta \right |  ,$$
so that after the changes of variables $g(x)$ 
has the form
\begin{eqnarray*}
g(x)= \int_0^1 \log\left| t-x\right| w(t) dt + \hbox{ regular part.}  
\end{eqnarray*}
For the convenience of some subsequent  proofs we shall also be
interested in  the iterated
equations obtained by replacing $f(t)$ under the integral 
by the right hand side of the integral equation $(\ref{e8})$:
\parbox{16cm}{\begin{eqnarray*}
f(x)& =& g_2 (x) +\lambda^2 \int_0^1  K_2(x,t)  f(t) dt ,\\
f(x)& =& g_3 (x) +\lambda^3 \int_0^1  K_3(x,t)  f(t) dt  ,\ \ldots
\end{eqnarray*}}\hfill \parbox{1cm}{\begin{eqnarray}\label{e10}
\end{eqnarray}}

\noindent and so on, where

\noindent \parbox{16cm}{\begin{eqnarray*}
g_2 (x)& =&g(x)+\lambda\int_0^1  K(x,t)   g(t) dt  , \\
g_3 (x)& =&g_2 (x)+\lambda^2\int_0^1  K_2(x,t) g(t) dt , \ \ldots
\end{eqnarray*}}\hfill \parbox{1cm}{\begin{eqnarray}\label{e11}
\end{eqnarray}}

\noindent and 

\noindent\parbox{16cm}{\begin{eqnarray*}
K_2(x,t)& =&\int_0^1  K(x,\tau )  K( \tau ,t) d\tau  ,\\
K_3(x,t)& =&\int_0^1  K_2(x,\tau )  K( \tau ,t) d\tau  , \ \ldots 
\end{eqnarray*}}\hfill \parbox{1cm}{\begin{eqnarray}\label{e12}
\end{eqnarray}}

\noindent If this iteration had  been continued indefinitely 
we would have found the Neumann series for $f$. Since these series 
usually converge only for very small values of $\lambda$, we shall 
not use them  but stop after a finite number of terms and 
take advantage of the fact  that the eigenvalues of 
$K_j (x,t)$ are the powers $ \{ \lambda_n^j \}$  
of  the eigenvalues $ \{ \lambda_n \}$ of $K(x,t)$. 
Indeed this will be quite helpful in some subsequent 
convergence proofs.

\vspace{0.3cm}
\noindent {\bsf IV. \qqr SINGULARITY OF THE FREE TERM}

In this and in the next section we shall try  to find 
the analytic structure of the edge
singularities of the solution  without solving the integral
equation, the latter  being possible only  numerically  or in 
some very special cases.${}^4$ To this aim we shall use methods similar 
to those from  the theory of the pinch or of the end point 
singularities,${}^{7}$ well known to particle physicists working 
in analytic S--matrix theory. As a preparation to what follows 
it is probably helpful to look to the corresponding 
chapters from the  classical book of Eden,  
Landshoff, Olive and Polkinghorne.${}^{7}$
 
In this section we shall deal with the free term 
$g(x)$ of Eq. $(\ref{e8})$. As mentioned in the Introduction, in order 
to find the analytic structure of the singularities we have to step 
into the neighbouring complex plane.
We shall start our investigations with some negative 
values $z_0$ of $z$ for which 
$\log  \left|t  -    z \right|   =   \log  (t  -   z )$
since the integration variable $t$ is between $0$ and $1$. Having to
perform analytic continuations we prefer to handle holomorphic 
expressions ($\log  \left|t  -   z \right|$ {\it is not} a 
holomorphic function of $z$) and so, instead working with $g(x)$
we shall focus our attention on functions of the kind
\begin{eqnarray}
F(z)= \int_0^1   \log  (t-z ) w (t) dt
\label{e13}
\end{eqnarray}
where the weight $w (t) $ is a function of $t$, which is 
holomorphic (no cuts  or other singularities) in neighbourhoods
of $t   =   0$ and $t  =   1$.
Although the holomorphic extension $F(z)$ is 
different from $g(x)$, it is closely related to it since,
up to regular terms, $\hbox {Re }F(x\pm i\varepsilon)  =   g(x)$ 
for $x  \in   [0,1]$ and  $\varepsilon \down 0$.
Since the weight $w(t)$ may differ very much 
from one case to another, integral (\ref{e13}) cannot
be performed explicitly. Therefore it will be interesting to have 
a mathematical procedure which should be able to predict the form 
of the singularity {\it without} actually performing the integrals. 
However, in order to have a partial check of the results which will be 
obtained below, we note that in the simplest case 
$w   \equiv   1$ we obtain $F_{w\equiv 1} 
(z)  =   z \log (-z)   +
  (1  -   z) \log (1  -   z)   -  1$.

As mentioned at the end of the last section, if 
$\lambda$ is small enough the Neumann series converge and so the 
solution can be written in terms of the free terms $g_j$ of the 
iterated equations. Therefore at the end of this section we shall 
discuss briefly  the singularity of the iterated functions $g_j (x)$
since they provide a check of the results obtained in Section  V 
in the general case.

\vspace{0.3cm}
\noindent {\bsf A. Different ways of defining a cut}

We recall that the features  which are important when considering
the cut structure of complex functions are the locations of
the branch points and  not the way in which the cut is taken. 
Indeed, the cut can be deformed or trailed as will become apparent 
below, in Figs. 3 to 5. As a first example, consider the function
$F_{w\equiv 1} (z) $ given above. To define the cut of the 
first term $z\log(-z)$, we first introduce the function 
$Z(z)  =  -z$ and ask then that the cut of 
$\log Z(z)$ should run along the positive 
$Z$ real axis. One achieves this by writing 
$Z   =  \left|Z\right|\exp (i\phi)$ 
and requiring $\phi$ to be in the range  $[ 0, 2 \pi)$.
With these conditions the cut of the first term, $z\log(-z)$,
of $F_{w\equiv 1} (z) $ runs along the negative $z$ real axis.
The function $z\log(-z)$ is real and equal to  $z\log \left|-z 
\right|   \equiv   -|z|\log\left|z \right|$
below the cut. Above the cut it will contain an 
additional imaginary part equal to $2i\pi z$.
There will be no cut along the positive $z$ real axis, 
here the value of $z\log(-z)$ is equal to  $z\log\left|z 
\right|  +   i\pi z$ both above and  below  the real axis.

It is not compulsory to take the cut of the second term
of $F_{w\equiv 1} (z) $ to the left. Indeed, if we define a new 
variable $Z(z)  =  1   -  z$, 
we can redefine the 'fundamental Riemann sheet' of  
$\log Z$ by requiring that the argument $\phi$ of $Z$ to be between 
$-\pi$ and $\pi$ rather than between $0$ and $2\pi$, so that the cut 
of $\log Z   =   \log (1   -  z)$ 
will run along the negative real $Z$ axis. Summarising 
$F_{w\equiv 1} (z)$ has no cut on the interval $[0,1]$ but 
two cuts running from $-\infty$ to $0$ and from $1$ to $\infty$. 

\vskip 0.3cm
\centerline{\epsffile [0 0 307 158] {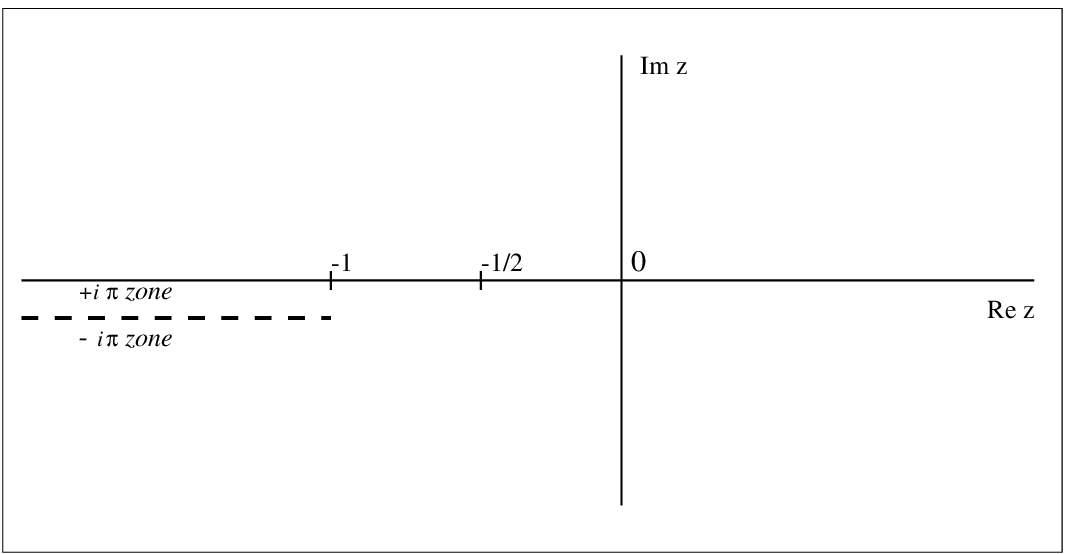}}

\centerline{{Fig. 2a.}}
\vskip 0.3cm 
\centerline{\epsffile [0 0 307 158] {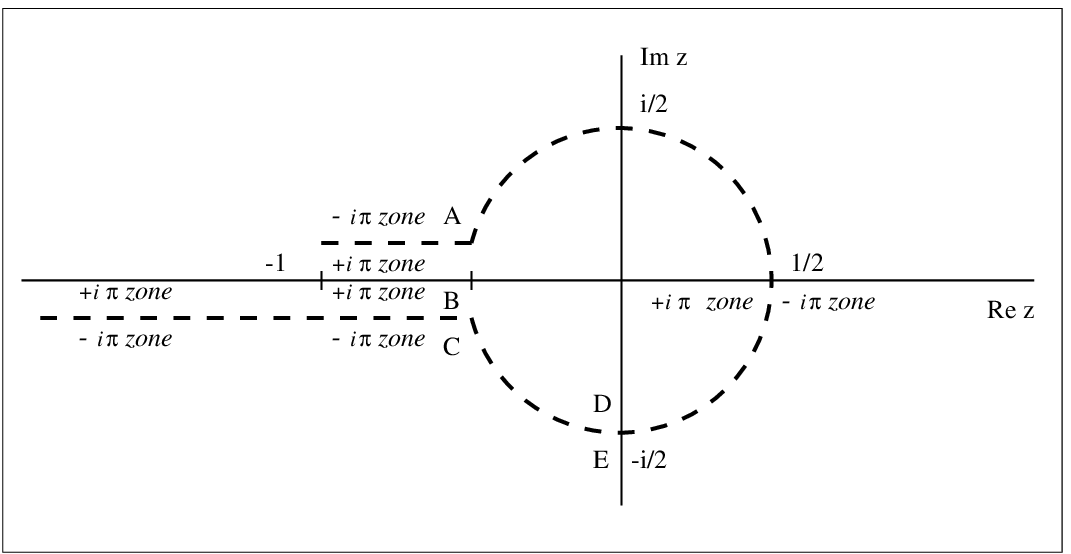}}

\centerline{{Fig. 2b.}}
\centerline{FIGS. 2. {\small Example of two different
definitions of the cut of $\log (z+1)$.}}

The possible patterns for the cut of the logarithm are not exhausted
by the cases discussed above. To have a further example let us
consider for instance the function 
$Z(z)  \equiv   1+z$ and begin with a cut of 
$\log Z$ running along the negative $Z$ real axis. 
In this case the cut of $\log (z+1)$ looks as in 
Fig. 2a where the $\pm i \pi \ zones$ mean that the values just above   
or beneath the cut differ by $\pm i \pi  $ from the mean 
value across the cut. (For illustration purposes  we have slightly
deplaced the cut: as it stands it corresponds to the function
$\log[z   -  (-1-i\varepsilon)]$, $\varepsilon >0$.)

However we could alternatively define the cuts to run as 
in Fig. 2b where again the $\pm i \pi \ zones$ mean that
the values of the logarithm differ by $\pm i \pi  $  
from the mean value across the cut (which  is not necessarily  real).
With these specifications one finds immediately that the value 
of $\log (z+1)$ at the points $A$,  $B$ and $C$ are, up to epsilons,
equal respectively to $\log \left|z_A +1 \right|$, to $\log
\left|z_A +1 \right| +2i \pi $ and again to $\log \left|z_A +
1 \right|$. As a further example,
the values of $\log (z+1)$ at the points 
$D(z  =  -i/2 + i\varepsilon$)
and $E(z  =  -i/2   -  i\varepsilon$) 
are  respectively
$\log \sqrt {5/4}  -  {(26.5/180)} i\pi   +   2 i \pi$ and
$\log \sqrt {5/4}  - {(26.5/180)} i\pi$, 
while  the mean value across the cut is there equal to 
$\log \sqrt {5/4}  - {(26.5/180)} i\pi   +   i \pi$.

\vspace{0.3cm}
\noindent {\bsf B. 'Hooking' the integration contour}

Let us come back to the function $F(z)$. In what follows it is
important to consider separately the parameter (the 'control') 
$z$--plane, and the complex $t$--plane where the integration 
is performed. As stated above, our aim is to find the 
singularities of $F(z)$, in the 'control' $z$--plane, 
without performing the integration explicitly. Since our final goal is
the description of the singularity  of $F(z)$ at $z=0$,
in this section we shall consider only analytic continuations performed 
in some neighbourhood of the origin. A similar discussion can be made
for the other end point $z=1$.

 \vskip 0.3cm 
\centerline{\epsffile [0 0 260 111] {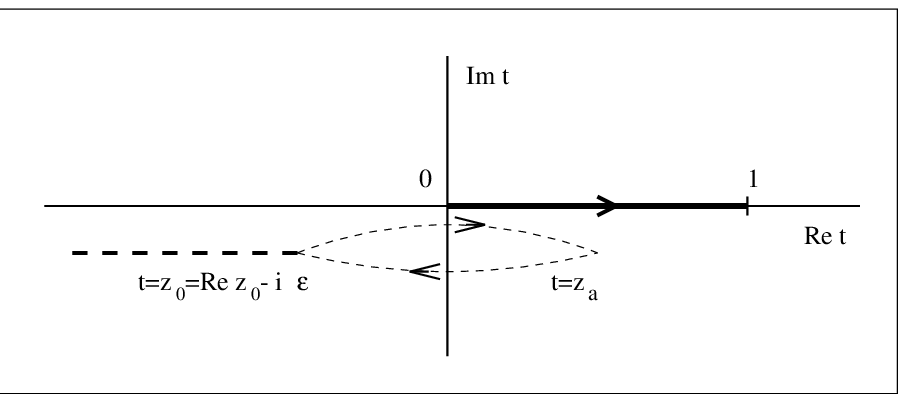}}
\centerline{FIG. 3. {\small Integration path (full line) and cut
of the logarithm (dashed).}}

Suppose that initially $z$ lies somewhere
immediately below the negative real axis in the complex $z$--plane: 
$z  =  z_0   \equiv   
x_0   -  i \varepsilon$, where
$x_0   <   0$ and $\varepsilon   >  0$. The function
$\log(t  -  z)$ from the integral of Eq. $(\ref{e13})$ has, 
{\it as a function of $t$} (i.e. in the complex $t$--plane where 
the integration is performed), a cut running parallel 
to the real $t$--axis from $t  =  -\infty$ to 
$t  =  z$  (see the dashed 
line in Fig. 3). From the point of view of the integration $t$--space, 
$z$ is a parameter. Suppose now that $z$ moves towards a point 
$z_a  =  x_a   -  i \varepsilon$ ($x_a   >  0$) 
and then returns to $z_0$ ($x_0   <   0$) without crossing the 
integration contour. Correspondingly, the head of the cut of the 
logarithm (as a function of $t$) will move in the $t$--plane as shown 
in Fig. 3, but, since  it will never cross the integration contour,
the value of $F (z_0 )$ will be identical with the value 
that the function $F (z)$ had before the point $z$ began 
to move from $z_0$.

\vskip 0.3cm 
\centerline{\epsffile [0 0 477 153] {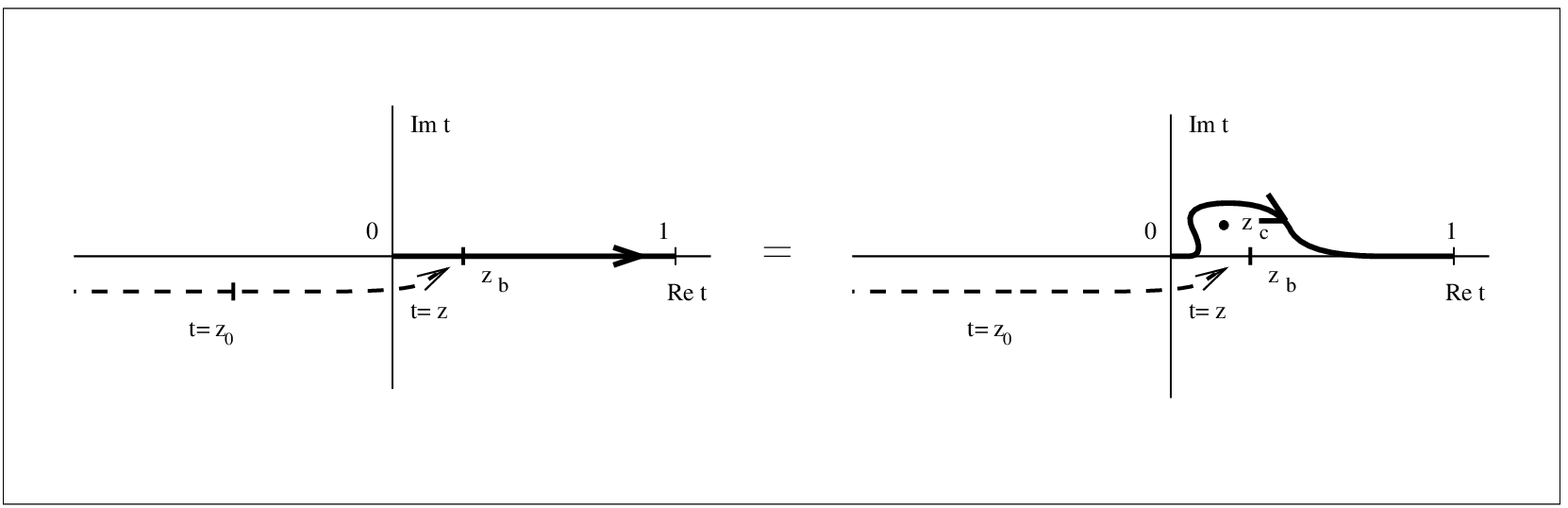}}
\centerline{FIG. 4. {\small The integrals over $[0,1]$ and 
over the deformed path are identical if $w(t)$ is holomorphic.}}

The situation is however  different if the
path followed by the point $z$ in the control complex $z$--plane
crosses the segment $[0,1]$ before returning to the initial
position $z   =   z_0$. Here again the path 
followed by the 'head' of the cut begins at $z_0$ and ends at the 
same point,  but this time it winds around the integration end point 
$t  =   0$, crossing the real axis at $t  =   z_b$ 
as shown in Figs. 4 and 5. 

\vskip .3cm
\centerline{\epsffile [0 0  315 228] {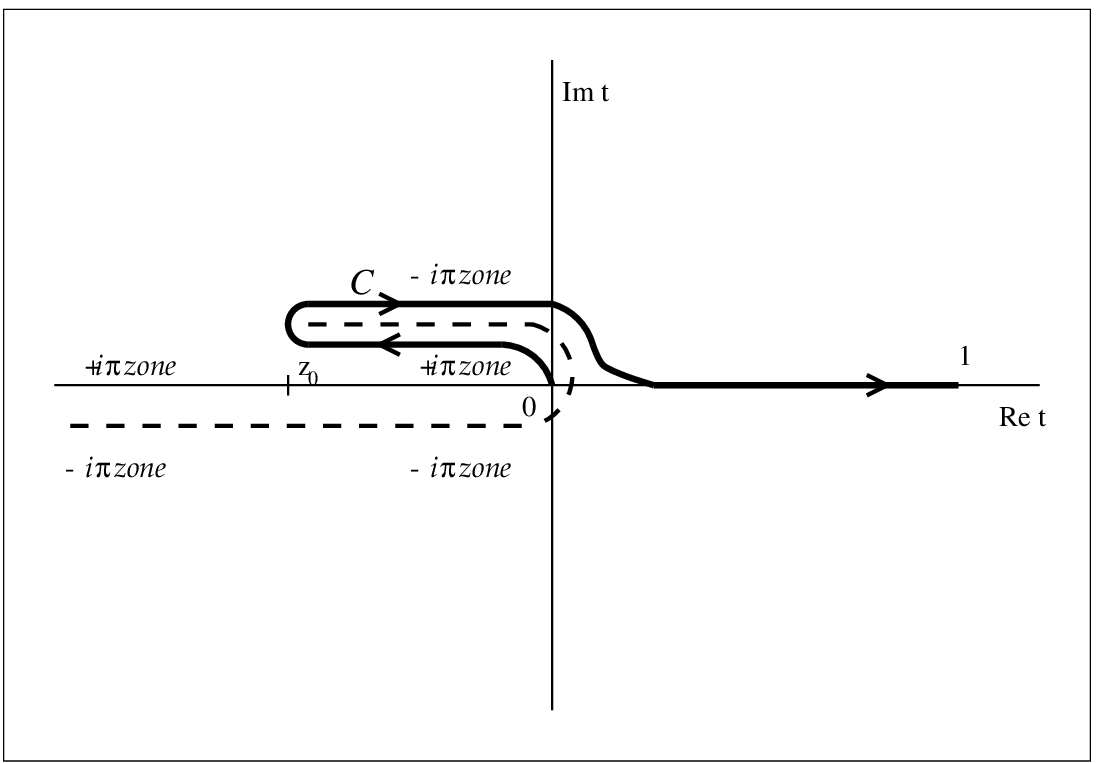}}
\centerline{FIG.  5. {\small The moving cut deforms the integration
 contour and  produces an Anomalous Threshold.}}

A genuine analytic continuation of the function $F(z)$ should, of
course, be at least continuous, i.e. have no jumps or other
discontinuities. Therefore, even before
the end point $t   =  z$ of the singularity of the logarithm 
reaches the point $t  =  z_b$ which lies on the real axis 
between $0$ and $1$, (see Fig. 4),
we shall use the freedom we have to deform the integration contours 
inside the analyticity domain of the integrand without altering in 
any way the value of the integral $F(z)$.

Since the branch point $t  =   z$ 'trails' behind 
it the cut of the logarithm  when $z$ moves further through points
$z_c$ in the upper half $z$--plane (Fig. 4) towards $z_0$, the value
we obtain for the analytic continuation $F^{(1)} (z_0)$ of the
initial integral will be  given by the integral along the
path {\it C} in Fig. 5 (the dashed line in Fig. 5
represents the cut of the logarithm; as it stands, the cut ends
at the conjugate point 
$\overline{z}_0   =  x_0   +  i\varepsilon$
rather than at $z_0$). Since  the value of the logarithm on 
the lower lip of the emerging part of the cut contains 
an additional $2\pi i$  with respect to that on the upper lip 
(recall the discussions about the points $A$ and $B$
from Fig. 2b), the  integral on the  part of the contour 
around the cut (see Fig. 5) is 
\begin{eqnarray*}
\int_{0}^{z_0}\bigl( \log \left| t-z_0 \right| + 2i \pi \bigr) w (t)dt
+\int_{z_0}^{0}  \log \left| t-z_0 \right|  w (t)dt
= - 2i \pi  \int_{z_0}^{0}  w (t)dt . 
\end{eqnarray*}
Here we have supposed that the weight $w (t)$ has no singularities
at $z  =  0$. Such a point $z_0$  where
the integration starts and which is below the initial
integration threshold is called an 'anomalous threshold'
in elementary particle physics.

The new value $ F^{(1)} (z_0)$ of $F(z)$  obtained by means of 
this analytic continuation process is hence 
\begin{eqnarray}  
F^{(1)} (z_0) = 2i \pi  \int_{0}^{z_0}  w (t) dt +
\int_{0}^{1}  \log ( t -z_0 )w (t) dt  
\label{e14}
\end{eqnarray} 
where the logarithm has the same determination as in Eq.  $(\ref{e13})$
(i.e. a cut like in Fig. 3).

There are many ways of defining the Riemann sheets of $F(z)$ and here
we describe only two of them: 

 (i) If  we now {\it define} $F (x+iy)$ for any point in the 
upper half $z$--complex plane to coincide with the function 
$F^{(1)}(x+iy)$ defined above, i.e. 
$F(x+iy)   \equiv     F^{(1)}(x+iy)$ ($y>0$),
we have implicitly required that the function
$F(z)$ should have no cuts on the segment $[0,1]$ but only on other
parts of the real axis. For instance, from  the above discussion 
it follows that along the negative real axis  $F(z)$ 
will have a discontinuity 
\begin{eqnarray}  
\Delta F(x) &=&F (x+i\varepsilon) - F (x-i\varepsilon)
\equiv F^{(1)} (x+i\varepsilon) - F (x-i\varepsilon) \nonumber\\
&= &2i \pi  \int_{0}^{x}  w (t) dt , \  
 x\in \hbox{negative real axis}, 
\label{e15}
\end{eqnarray} 
which means that $F(z)$ has indeed a cut along the negative real axis.
This definition of the 'fundamental' Riemann sheet of $F(z)$ coincides 
with that for the simple example of $F_{w\equiv 1} (z)$ discussed 
in the previous subsection.

 (ii) Alternatively,  we could have required that the function 
$F(z)$ should have no cuts along the negative real axis. This amounts 
to redefining its 'fundamental' Riemann sheet:  we start again from our 
$z_0   \equiv   x_0   -  i \varepsilon$ with $x_0  < 0$,
 ask that 
$F^{(0)}(z_0)    \equiv   F(z_0)$ but then require 
that the values of $F^{(0)}(z)$ above the real axis should merge, 
for $\hbox {Re } z   <   0$, with those 
below the axis. We, therefore, {\it define} $F^{(0)}(z)$ for
 $\hbox {Re } z   <   0$ as
\begin{eqnarray} 
F^{(0)}(z)= \int_0^1   \log  (t-z ) w (t) dt  .
\label{e16}
\end{eqnarray}
This definition which initially has been given only for 
$\hbox {Re }z   <   0 $ may then be extended to the whole complex 
plane cut along the real segment $[0,1]$ (and further, along the
positive real axis up to infinity; but, as mentioned at the beginning
of this subsection, in order to keep the discussion simple we shall not
consider whet happens beyond the end point $z=1$).
 According to the definition 
 (ii), the function $F^{(0)}(z)$ will 
have no cuts along the real negative axis, but in return, will have  
different values along the upper and the lower lip of the real 
segment $[0,1]$. In the lower half plane $F^{(0)} (z)$ coincides 
with the function   $F (z)$ corresponding to the 
previous definition (i) of the fundamental Riemann sheet,
 but not any longer for $\hbox {Im } z  >   0$ where $F (z)$ was
identical to $F^{(1)} (z)$. However, since by the construction
of the function 
$F^{(1)} (z)$ $[$see the discussion preceding Eq. $(\ref{e14})$$]$
 we had 
\begin{eqnarray} 
F^{(0)}(x-i \varepsilon)= F^{(1)}(x+i \varepsilon) +{\cal O} 
(\varepsilon)  , \hbox{ for } \varepsilon > 0 \hbox{ and }  0<x<1  ,
\label{e17}
\end{eqnarray}
it follows that $F^{(1)}(z)$ represents now the analytic
continuation of  $F^{(0)}(z)$ on the next Riemann 
sheet---call it sheet $(1)$---when one 
crosses the segment $[0,1]$ in the {\it upward} direction.

On the other hand, starting from the values $ F^{(0)} (x+i\varepsilon)$
from the upper lip of the segment $[0,1]$ and continuing them
{\it downwards}, one gets the function $F^{(-1)}(z)$: 
\begin{eqnarray} 
F^{(-1)} (x-i\varepsilon)=F^{(0)} (x+i\varepsilon)
+{\cal O} (\varepsilon) , \hbox{ for } \varepsilon > 0 \hbox{ and } 
  0<x<1 
\label{e18}
\end{eqnarray}
which is obtained by deforming the integration contour 
in the lower half plane. Hence, $F^{(-1)}(z)$ will have the form
\begin{eqnarray}  
F^{(-1)} (z_0) = -2i \pi  \int_{0}^{z_0}  w (t) dt +
\int_{0}^{1}  \log ( t -z_0 )w (t) dt 
\label{e19}
\end{eqnarray} 
where, again, the logarithm has the same determination as in Eqs. 
$(\ref{e13})$ and $(\ref{e16})$.
Further, we see that the jump of $F^{(0)}$ across
the cut $[0,1]$ is 
\begin{eqnarray*}
\Delta F^{(0)} (x) = -2i \pi  \int_{0}^{x}  w (t) dt  , \qqr
\hbox{ for }  0<x<1  . 
\end{eqnarray*}
The functions $F^{(-1)}$, $F^{(0)}$, $F^{(1)}$ 
living on the Riemann sheets $(-1)$, $(0)$, $(1)$ represent in fact
an unique analytic function  $"F(z)"$, the various branches 
$F^{(-1)}(z)$, $F^{(0)}(z)$, $F^{(1)}(z)$ being nothing but 
its values on a cutting up of the initial Riemann manifold 
along some arbitrarily given cuts.

If the function $w$ is holomorphic in some neighbourhood of the 
origin it admits there an expansion of the form
\begin{eqnarray}  
w (t )= a_0 +  a_1 t +a_2 t^2  + \cdots   ,
\label{e20}
\end{eqnarray} 
and so, from Eq. $(\ref{e15})$ we find 
\begin{eqnarray}  
\Delta  F(x) =  2i \pi \Bigl[ a_0 x +{a_1\over 2}x^2  
+{a_2\over 3}  x^3 + \cdots \Bigr] . 
\label{e21}
\end{eqnarray} 
A function which has precisely the same jump along the negative axis 
is given by 
\begin{eqnarray} 
\Bigl[ a_0 +  {a_1 \over 2} z+ 
{a_2 \over 3}z^2 + \cdots \Bigr] z \log z
\label{e22}
\end{eqnarray} 
which provides a mathematically correct and at the  same time
extremely simple description of the singularity of $F(z) $
near the point $z  =  0$. 

Before closing this section we shall
discuss briefly  the structure of the singularities at
the origin  of the free terms $g_2(z),\ g_3(z),\  \ldots ,\ g_j(z)$
of the iterated equations $(\ref{e10})$. In contrast to $g(z)$, these
functions are written as integrals of the form $(\ref{e13})$ over
weights $w(t)$ which are no longer holomorphic but contain 
the singularities of $g(z),\ g_2(z),\ \ldots ,\ g_{j-1}(z)$ 
respectively. By straightforward integration we find that 
the general term of the iterated function $g_j(z)$ is 
\begin{equation}
z^m \log^k(z), \qqr\hbox{ with }k=0,1,...,j \hbox{ and } m\ge k  .
\label{e23}
\end{equation}

At this stage we may wonder whether the left hand cuts of the
powers of $\log(t)$ appearing under the integrals will not
interfere with  the above analytic continuation process.
This does not occur since the real part of $t$ becomes negative 
only along the loop in Fig. 5 around the emerging part of the
cut of $\log (t  -   z)$ which is always at the same side of 
the negative real axis, i.e. always in the same '$+i\pi\ zone$' of 
the function $\log(t)$. This means that
\begin{eqnarray}
\Delta \left[ \int_0^1    \log  (t-z ) \ t^m \log^k(t) dt \right] =
2i\pi \int_{0}^{z}  t^m(\log |t| +i\pi)^k dt,
\  z\in \hbox{negative real axis} .
\label{e24}
\end{eqnarray}

\noindent {\bsf V. \qqr SINGULARITIES OF THE SOLUTION OF THE FULL 
 EQUATION}

 As  mentioned in  the Introduction, the specific difficulties
of our problem are twofold. We have first to handle moving cuts; 
this question has been largely discussed in the previous section 
devoted to the free term. Second---and this is probably  the
main difference with  respect to the classical papers
on pinch and end point singularities---we will have 
to cope with the  fact that the singularities of
the function under the integral are {\it a priori} unknown,
this function being  the solution of the integral
equation whose analytic properties we are trying to establish.
In this section we shall address this second problem by
solving it first for the eigenfunctions which are the natural
building blocks of the solution, with the hope that their
analytic properties (together with those of the free term) will be
transmitted to the solution itself. Of course, this is not at all 
obvious since we will deal with infinite series and so new singularities
may creep in. Therefore before we embark  in Section  V B on  
the study of the Riemann sheet structure and the asymptotic expansion
of the  eigenfunctions, we will first make sure in subsection
V A that the analytic properties of the eigenfunctions do
exhaust the holomorphic characteristics of the solution.
This is probably not entirely pedagogical, but reflects fairly well
the way in which our work  progressed. 
We  shall have often to refer in Section
 V A to various analytic properties of the eigenfunctions
which will  be proved only later in Section  V B. 

This type of analysis presented here is not restricted to this 
particular integral equation, but we hope that it provides a working 
example of how one could proceed in problems involving other
kinds of weakly singular kernels. To this end we tried to avoid as
much as possible any special properties of the logarithmic
kernel---for instance the fact that its null--space 
$ \hbox{ker \bsf K}$  is empty---and show how we can
proceed in the general case.
 At a first reading or if  not particularly interested
in these mathematical proofs but only in the practical aspects of
the asymptotic expansion, the reader may read only Section 
 V A 1, skip the remainder of the Section  V A and
pass directly to Section  V B.

\vspace{0.3cm}
\noindent {\bsf A. \qqr The absence  of  summation singularities}

We shall proceed in a number of stages: 

In Section $ V \ A \ 1$ we discuss the role of  eigenfunction
expansions in describing the generic singularities of the 
solution. In both  Sections $ V \ A \ 1$ and $ V \ A \ 2$ we recall 
some facts from the theory of integral equations and we identify a 
class of functions $\{ \psi \}$ which can be expanded in terms of
the eigenfunctions. We discuss also the possible appearance of
additional  singularities due to problems of convergence
of infinite sums of functions.

The convergence proofs are provided in two separate
subsections. In Section  $ V \ A \ 3$ we discuss the continuity of
these $\psi$--functions on the real segment $[0,1]$ and 
we show that their eigenfunction expansions converge
uniformly there. However, in order to be able to consider 
the holomorphic properties of the  solution $f(z)$ we need a number
of results in the complex $z$--plane. These are derived in 
Section  $ V \ A \ 4$ where, in particular, we prove that no
additional singularities appear as a result of the summation 
of the series. The convergence of the asymptotic parts of
the eigenfunctions is discussed in $ V \ A \ 5$.
 
\vspace{0.3cm}
\noindent{\bsfl 1.  Generic singularities and
eigenfunction expansions}

 When discussing the possible singularities of the solution of
an integral equation we are  faced with an apparent paradox. 
Independently of the specific form of the integral kernel,
we may always proceed as a numerical analyst usually does when 
checking the correctness of
computer code:   start in the right hand  side of
Eq. $(\ref{e8})$ with some 'nice' function $f(t)$ which has no 
singularities, integrate it with the kernel and then choose the free
term $g(x)$  to recover the  initial function $f(x)$.
So, irrespective of the (integrable) singularities of the kernel, 
the solution $f(x)$ might be a polynomial, a simple trigonometric 
function or anything else. One may feel that this type of solution 
is quite exceptional but this example is enough to show that one 
{\it cannot} speak  about 'the general singularity' of the solution of
an integral equation with a given singular kernel. However one is
fully entitled to ask oneself what may happen in the 
non--exceptional situations, i.e. in the {\it generic} case.

To this end it is enlightening to express the solution of 
the integral equation $(\ref{e8})$ for our logarithmic kernel 
in terms of the eigenfunctions $u_n (x)$ of $K(x,t)$, defined by 
\beq
u_n (x)= \lambda_n \int_0^1  K(x,t)  u_n (t) dt , \qqqr 
\hbox{with }K(x,t) \equiv \log \left|t- x\right| ,
\label{e25}
\eeq
as the series 
\beq
f(x) = g(x) +\sum_{n=0}^{\infty} {\lambda \over \lambda_n - \lambda}
g_n u_n(x) ,
\label{e26}
\eeq
derived below in Section  $ V \ A \ 2$, where
\beq
g_n  \df \int_0^1 g(x) u_n(x) dx .
\label{e27}
\eeq
From expansion $(\ref{e26})$ it is obvious that in the 
{\it generic} case when small changes in the form of the 
function $g(x)$ and hence in $\{ g_n \}$ are allowed,
both the singularities of the free term  $g(x)$ and of the
eigenfunctions $u_n (x)$  will appear in the solution $f(x)$  
since they will no longer cancel identically.

\vspace{0.3cm}
\noindent {\bsfl 2.  The functions} \mbox{\boldmath$\psi (x)$}

In what follows  we shall use systematically the notation
$\psi (x)$  for the functions from the range 
$ \hbox {Ran \bsf K}$ 
of the integral operator. The properties of these functions 
are used in the derivation of expansion 
$(\ref{e26})$ which plays a central role for the analytic
properties of $f$ as a superposition of those of $g$ and
of the $u_n$. However we should exercise great care at this
point since additional singularities may creep 
in. We should have in mind the case of the sequence of functions
$1, 1+z, 1+z+z^2, \ldots \ $. 
In the open unit disk
this sequence tends to the function $1/(1  -  z)$
which has a pole at $z  =  1$, whereas  
all the functions  in the sequence are  holomorphic in an arbitrary 
large disk. Later in  Section  $ V \ A \ 4$ we will use a 
theorem of Vitali and/or of Morera to prove that no additional 
singularities appear in a neighbourhood of the origin. The theorem 
of Vitali, for instance, is partially based on the  requirement that 
the elements of the sequence of partial sums should be uniformly 
bounded, which is clearly not valid in the counter-example with 
the geometric series if $|z|   \ge   1$. Hence we have to 
make sure that in our case all the requirements of these theorems 
are fulfilled.

In order to obtain Eq. $(\ref{e26})$ we first
multiply the integral equation $(\ref{e8})$ by an
eigenfunction $u_n(x)$, integrate over the variable $x$
and  use the symmetry of the kernel to get
\beq
f_n=g_n+{\lambda \over \lambda_n} f_n  , 
\label{e28}
\eeq
where the coefficients $f_n$ are defined from $f(x)$ by integrals 
similar to those in Eq. $(\ref{e27})$.

From Eq. $(\ref{e28})$ we find 
$f_n   =  \lambda_n g_n /(\lambda_n  -
  \lambda)$, 
but we should avoid expressing the solution $f(x)$ as a sum
$\sum f_n u_n(x)$ since the latter might not converge pointwise
and/or the eigenfunctions $\{ u_n \}$ might not represent a 
complete system of orthonormal functions. In the special case of
the logarithmic kernel $(\ref{e25})$ it happens (see Appendix) that
the $\{ u_n \}$ do represent such a complete orthonormal system, 
but for an arbitrary kernel $K$,
$\hbox{ker \bsf K}$ is not empty and so they do not. 

Many of the textbook theorems concerning expansions of the type 
$(\ref{e26})$ rely on the continuity of the kernel in the square 
$[0,1] \times [0,1]$. Since this  is certainly not the case for
our logarithmic kernel,  some supplementary work is necessary
to adapt the proofs to our specific conditions.
In the simple cases where the kernel is continuous one 
usually takes  advantage of this fact  to prove
that for any square integrable---even singular---function
$\varphi(t)$, the  function  
\beq
\psi(x) \df \int_0^1 K(x,t) \varphi(t) dt 
\label{e29}
\eeq
from $ \hbox {Ran }\hbox{\bsf K}$ is (a) continuous for $x\in [0,1]$, 
and (b) expressible in the form of an uniformly convergent 
expansion
\begin{eqnarray}
\psi(x)&= &\sum_{n=0}^{\infty} \psi_n u_n(x) ,\label{e30}\\
       &\equiv & \sum_{n=0}^{\infty} \varphi_n/\lambda_n u_n(x) 
       \nonumber 
\end{eqnarray}
where $\psi_n$ and $\varphi_n$ are defined by
\begin{eqnarray*}
\psi_n=\int_0^1 \psi(t) u_n(t) dt , \ \ 
\varphi_n=\int_0^1 \varphi(t) u_n(t) dt  .
\end{eqnarray*}
Here the relation which connects the coefficients  $\psi_n$ and 
$\varphi_n$ can be obtained by multiplying Eq. $(\ref{e29})$ by 
$u_n (x)$, integrating and using the symmetry of the kernel  
\beq 
\psi_n = \varphi_n/\lambda_n . \label{e31}
\eeq
The solution $f(x)$ itself does not in general have a
representation of the $\psi$--kind $(\ref{e29})$, but  it is clear 
from the integral equation $(\ref{e8})$ that the difference 
$f(x)  -   g(x)$ is a true $\psi$--kind function. 
Hence it can be expanded as the sum
$\sum (f_n   -   g_n) u_n(x)$, 
which in turn, using the relation $(\ref{e28})$
between $f_n$ and $g_n$, yields the representation $(\ref{e26})$
for the solution of the integral equation in terms of the free
term $g(x)$ and of the eigenfunctions $u_n(x)$.

In  Section $ V \ A \ 3$ we shall prove that the properties  (a)
and  (b) and hence the expandibility of  
$f(x)  -   g(x)$ are also valid in the case of the logarithmic 
kernel. Before showing that let us notice that if $\hbox{ker \bsf K}$ 
is empty as is the case---see the Appendix---with the logarithmic
 kernel, or if $\varphi$ is chosen from the orthogonal complement 
$\hbox{ker}^\bot\, \hbox{\bsf K}$ of $\hbox{ker \bsf K}$, we also have
\beq 
\left\|\varphi(x)-\sum_{n=0}^{\infty} \varphi_n u_n(x)  
\right\| _{L^2} =0  .
\label{e32}
\eeq
However, in contrast with what happens with the $\psi$--kind
functions, Eq.  $(\ref{e32})$ represents only a convergence in the mean,
i.e. $\varphi(x)$ does not have, in general, an expansion of
the form $(\ref{e30})$ which converges uniformly. 
For the study of the analytic properties of the 
solution we shall need finer properties 
than those offered by $L^2$--space arguments,
since, for instance, holomorphy and uniform convergence 
of the partial sums are essential for the Morera theorem to be
used in Section $ V \ A \ 4$. 

\vspace{0.3cm}
\noindent{\bsfl 3. Continuity of \mbox{\boldmath$\psi (x)$} in the 
logarithmic case and the uniform convergence of the \\
 \mbox{\boldmath$\psi^{(j)} (x)$} on the segment 
[0,1]}

This subsection deals with the properties of the $\psi$--kind
functions $(\ref{e29})$ on the segment $[0,1]$ for the 
logarithmic kernel  $(\ref{e9})$. The arguments are quite similar
to those which are used in the case of the continuous kernels,
but we shall discuss them briefly here as a preparation for the 
next subsection and as well as to make this paper self contained.

Our logarithmic 
kernel becomes infinite each  time the integration variable 
$t$ equals $x$. However, the continuity   (a) of $\psi (x)$ on 
the interval $[0,1]$ (including at its end points) can be proved in 
a straightforward manner using the Schwarz inequality.
Indeed, for any $L^2$ function $\| \varphi \|_{L^2} \leq M$ and
for any points $x$ and $x+\delta$ belonging to the (closed) segment
 $[0,1]$, we have
\beq
|\psi(x+\delta)-\psi(x)|^2 \leq \int_0^1 
[\log \left|t-(x+\delta )\right| - \log \left|t- x\right| ]^2 dt 
\times M^2 
\label{e33}
\eeq
where the integral $\int_0^1 \left| \varphi(t)\right|^2 dt $ has been
replaced by the bound on the $L^2$--norm. If $\delta$ is 
small enough, it can be shown that
the integral on the right hand side of $(\ref{e33})$ can 
be made smaller than any given nonzero $\varepsilon^2/M^2$, which
proves the continuity of  all the $\psi$--kind functions  
of the form $(\ref{e29})$. Since the eigenfunctions $u_n(x)$ by their 
very definition $(\ref{e25})$ are also functions of the  $\psi$--kind,
we have hence implicitly proved their continuity on
the segment $[0,1]$, including at the end points
$x   =   0$ and $x  =  1$.

We shall now show,  (b), that for $x\in [0,1]$ the finite sums
\beq
\psi^{(j)}(x)=\sum_{n=0}^j \psi_n u_n(x) 
\label{e34}
\eeq
converge uniformly  to
the function $\psi (x)$ defined in Eq. $(\ref{e29})$.

 To this end we shall show first that the functions 
$\psi^{(j)}(x)$ converge uniformly to some (continuous) function
$\psi^{(\infty )}(x)$:  a similar proof may be used then
for the uniform convergence of  the analytic extensions 
$\Psi^{(j)} (z)$ which will be introduced in the next subsection.
From Eq. $(\ref{e31})$ and from the definition $(\ref{e25})$ of the 
eigenfunctions $\{u_n (x)\}$  we have
\begin{eqnarray*}
\left| {\textstyle\sum_{j+1}^{j+k} } \psi_n u_n(x) \right|^2
\equiv \left|  {\textstyle\sum_{j+1}^{j+k} }
{\varphi_n \over \lambda_n} \cdot \lambda_n {\int_0^1 K(x,t)u_n(t)dt }  
\right|^2 = \left|\int_0^1 K(x,t) {\textstyle\sum_{j+1}^{j+k}}
\varphi_n u_n(t)dt\right|^2,
\end{eqnarray*}
which, using the Schwarz inequality and the fact that the 
basis $\{ u_n (t) \}$ is orthonormal, yields: 
\beq
\left| {\textstyle\sum_{j+1}^{j+k} } \psi_n u_n(x) \right|^2
\le \int_0^1 \left| K(x,t) \right|^2 dt \cdot \int_0^1 \left|
{\textstyle\sum_{j+1}^{j+k}}
 \varphi_n u_n(t)\right|^2 dt=\int_0^1 \left| K(x,t) \right|^2 dt \cdot
{\textstyle\sum_{j+1}^{j+k}}\varphi_n^2  .
\label{e35}
\eeq
For each fixed value of $x$, $0\le x\le 1$,
the kernel $K(x,t)\equiv \log |t-x|$ regarded as a function of $t$
is in $L^2 [0,1]$, and so,
\begin{eqnarray*}
\int_0^1 K^2 (x,t) dt \ <M^2 <\infty  .
\end{eqnarray*}
Since  the function $\varphi$ is also in  $L^2$,
${\textstyle\sum_{j+1}^{j+k}} \varphi_n^2$ tends to zero for 
$\forall k$ as $j$ 
increases. Hence  the right hand side of $(\ref{e35})$
can be made arbitrarily small 
irrespective of the value of  $x$.
This means that the sequence $\psi^{(j)}(x)$ converges {\it uniformly}
to some limit $ \psi^{(\infty )} (x)$. Now
from the continuity of the eigenfunctions  $\{ u_n (x)\}$ which
we have proved above, it follows that the finite combinations
$\{ \psi^{(j)} (x)\}$ are continuous. Since the $\{ \psi^{(j)} (x)\}$
converge {\it uniformly},  the limit $ \psi^{(\infty )} (x)$ is also
continuous.
 
On the other hand, by projecting the kernel onto the basis 
$\{ u_n  \}$ we obtain 'the coefficients' $ {u_n (x)/ \lambda_n} $.
Bessel's  inequality then  ensures that any sum over the
$u_n^2 (x) / \lambda_n^2$  is bounded: 
\begin{eqnarray}
\sum_{n=0}^{\infty} {u_n^2(x) \over \lambda_n^2} \le \int_0^1 K^2 (x,t)
dt \ (<\infty )  .
\label{e36}
\end{eqnarray}
Integrating with respect to $x$  we see that 
$\sum 1/ \lambda_n^2$ must also converge  and 
hence  the $| \lambda_n |$ must  tend to infinity with $n$. 
This fact will help us to prove that $\psi^{(\infty )}(x)$ is in
fact identical to the function $\psi(x)$ defined in Eq. $(\ref{e29})$. 

Indeed, since both functions $\psi(x)$ and $\psi^{(\infty)}(x)$ are 
continuous, it is enough to show that the $L^2$--norm of their 
difference is zero. To this end we notice that the difference between 
$ \psi (x)$ and the functions
$ \psi^{(j)} (x)$ can be written as  
\beq
\psi(x)-\psi^{(j)}(x)=\int_0^1  K^{(j+1)}(x,t) \varphi(t) dt
\label{e37}
\eeq
where $K^{(j+1)}(x,t)$ is the truncated kernel
\begin{eqnarray} 
K^{(j+1)}(x,t)=K(x,t)-\sum_{n=0}^j {u_n(x)u_n(t) 
\over \lambda_n}  ,
\label{e38}
\end{eqnarray}
where we have supposed that the eigenvalues have been 
labelled according their increasing \linebreak moduli: 
$\left| \lambda_0 \right| \le \left| \lambda_1 \right|  \le \cdots$.
Unlike the procedure followed before,
we shall not try to use the Schwarz inequality 
to prove directly the pointwise convergence of the $\psi^{(j)}(x)$,
but we shall go instead through $L^2$--space arguments. Since the
first $j  +  1$ eigenfunctions $u_0 (x),\ u_1 (x),\ \ldots,
\ u_j (x)$ are all in the  null space $\hbox{ker \bsf K}^{(j+1)}$ 
of the truncated kernel $(\ref{e38})$,
it follows that its eigenvalue with smallest absolute value 
is $\lambda_{j+1}$. If we denote by $ K^{(j+1)}_2(x,y)$
the iterated  truncated kernel
\begin{eqnarray} 
 K^{(j+1)}_2(x,y) = \int K^{(j+1)}(x,t) K^{(j+1)}(t,y) dt   ,
\label{e39} 
\end{eqnarray}
its smallest eigenvalue $\mu$ will be $\lambda^2_{j+1}$. However 
for any symmetric  Hilbert--Schmidt kernel ${\cal K} (x,y)$  we have 
$$ \sup_{\|\varphi\|=M} \left| \int \!\!\int {\cal K}(x,y)
\varphi(x)\varphi(y) dx dy \right| = {M^2 \over |\mu |}$$  
where $\mu$ is  the smallest eigenvalue of ${\cal K}$.
Hence, taking  the integral over the square of the modulus of the left 
hand side of Eq. $(\ref{e37})$ we obtain
\begin{eqnarray}
\|\psi-\psi^{(j)} \|_{L^2}^2=
\left| \int \!\!\int K^{(j+1)}_2(x,y) \varphi(x)\varphi(y)dx dy \right|
\le {M^2 \over \lambda_{j+1}^2}
\label{e40}
\end{eqnarray}
which means that $\|\psi-\psi^{(j)} \|_{L^2} 
\le {M /{\left| \lambda_{j+1}\right|}}$. \qqr
Since  $1/{\left| \lambda_{j+1}\right|}$ tends to zero as $j$ 
increases, so does $\|\psi-\psi^{(j)} \|_{L^2}$. Now, from the
uniform convergence 
$\|\psi^{(j)}-\psi^{(\infty)} \|_{L^{(\infty)}}\rightarrow 0$ 
which has proved above [see Eq. $(\ref{e35})$], we have immediately
also the $L^2$ convergence
\begin{eqnarray}
\|\psi^{(j)}-\psi^{(\infty)} \|_{L^2}\rightarrow 0  .
\label{e41}
\end{eqnarray} 
From the triangle inequality we have
\begin{eqnarray}
\|\psi-\psi^{(\infty)} \|_{L^2} \le \|\psi-\psi^{(j)} \|_{L^2} +
\|\psi^{(j)}-\psi^{(\infty)} \|_{L^2}  ,
\label{e42}
\end{eqnarray}
where from Eqs. $(\ref{e40})$ and $(\ref{e41})$ we see that the left 
hand side, which is independent of $j$, can be made arbitrarily 
small by a suitable choice of $j$ in the right hand side.
This means that $\|\psi-\psi^{(\infty)} \|_{L^2}\equiv 0$ which,
since both $\psi (x)$ and $\psi^{(\infty)} (x)$ are continuous,
proves that the two functions are identical.

\vspace{0.3cm}
\noindent {\bsfl 4.  The functions  \mbox{\boldmath$\Psi^{(j)}(z)$}
and the theorems of Morera and Vitali}

In order to be able to study the nature of the singularities
of the functions of $\psi$--kind,
we will have to step off the real axis into the complex 
$z$--plane. We shall be particularly interested in the complex plane 
singularities and the asymptotic expansions near $z  =  0$ 
of eigenfunction expansions of the form  $(\ref{e26})$,  
related to the solution $f(x)$ of the integral equation.
To this end we shall need to know some analytic properties
of the eigenfunctions $\{u_n\}$. These will be derived
in Section V B where, similarly to the 
function $F^{(0)} (z)$ [see Eq. $(\ref{e16})$ from Section  IV], 
we shall define the analytic functions
\beq
 U{}_n^{(0)} (z) \df  \lambda_n \int_0^1 \log (t-z) u_n (t) dt  , 
\qqr n=1,\ 2, \ 3,\ \ldots  .
\label{e43} 
\eeq
Although these functions  do not yet represent the analytic
continuation of the  eigenfunctions $u_n (x)$ which are defined
on the real segment $[0,1]$, they are closely related  to them.
This specific choice is based, as for $F^{(0)}(z)$, on the fact 
that for real negative $z$ we have $\log (t  -  z)
  \equiv   \log |t   -  z|$, since 
the integration variable $t$ on the right hand side of
Eq. $(\ref{e43})$ is always between 0 and 1. In contrast to the
function $\log |t   -  z|$ (which is {\it not}  holomorphic 
because of the modulus), the function $\log (t  -  z)$ can be
extended analytically in the whole (cut) complex $z$--plane.

As will become apparent in Section V B, the analytic functions
\qqr $ U_n^{(0)}(z)$, \qqr $ U_n^{(1)}(z)$,\qqr  $ U_n^{(-1)}(z)$,
$\ldots $ which are the analogues of the functions $F^{(0)}(z)$,
$F^{(1)}(z)$, $F^{(-1)}(z)$, $\ldots $, defined in Eqs. $(\ref{e16})$,
$(\ref{e14})$ and $(\ref{e19})$, have an infinite Riemann sheet
structure; the superscript in parantheses indicates the Riemann
sheet under consideration. From the  definition of the Riemann sheets
and the continuity properties across the cut, we have, similar to
 Eqs.$(\ref{e17})$--$(\ref{e18})$,  for  any $x$ real
 between $0$ and $1$: 
\beq
U_n^{(k)}(x+i\varepsilon)=U_n^{(k-1)}(x-i\varepsilon)+{\cal O} 
 (\varepsilon), \qqr \varepsilon>0, \qqr x\in [0,1] .
\label{e44}
\eeq
 As will be shown in Section $V\ B\ 3$, the eigenfunctions
$u_n(x)$  defined on the interval $[0,1]$ can be written
as simple combinations of the boundary values of 
$U_n^{(0)}(z)$, $U_n^{(1)}(z)$ and $U_n^{(-1)}(z)$
on the upper and lower lips of the cut:  
\begin{eqnarray}
&& u_n(x) = {1\over 2}(U_n^{(0)}(x+i\varepsilon)+
U_n^{(1)}(x+i\varepsilon)),\qqr 
 \varepsilon \down 0,\qqr 0<x<1  ,  
\nonumber \\
&& u_n(x) = {1\over 2}(U_n^{(0)}(x-i\varepsilon)+
U_n^{(-1)}(x-i\varepsilon)),\qqr \varepsilon \down 0,\qqr 0<x<1  .
\label{e45}
\end{eqnarray}

Hence, in analogy with the function $\psi (x)$ 
and the finite sums $\psi^{(j)} (x)$ defined in the previous
subsection for $x\in [0,1]$, we shall define now
for $z$ in some given region $\Omega$ of the
complex plane, the holomorphic functions 
\beqa
&&\Psi(z)\df \int_0^1 \log(t-z) \varphi(t) dt  ,
\qqr \varphi\in L^2[0,1]  , \label{e46} \\
&&\Psi^{(j)} (z) \df \sum_{n=0}^j \psi_n U_n^{(0)}(z) \id 
\sum_{n=0}^j \frac{\varphi_n}{\lambda_n} U_n^{(0)}(z) .
\label{e47}
\eeqa
Since we are mainly interested in the behaviour of the solution
near the origin, it is sufficient to take the region $\Omega$ to be
the (open) unit disk cut along the real segment $[0,1]$ (see Fig. 6),
but any other (cut) disk of radius $R$ is also acceptable. The 
holomorphy of $\Psi(z)$ in $\Omega$ follows 
from the  theorem of Morera  which states that 
iff ${}^{11,12}$ the function  $\Psi(z)$ {\it is continuous}
in the open set $\Omega$  and iff
\beq 
\int_{\partial \Delta} \Psi (z) dz = 0
\label{e48}
\eeq 
along the border of {\it every} closed triangle $\Delta \subset \Omega$,
(i.e. along any reasonable regular closed contour), then the function
$\Psi(z)$ {\it is holomorphic} in $\Omega$. It is clear that 
these two conditions are met by any  function having the
representation $(\ref{e46})$ for $z   \in   \Omega$. 
The continuity of $\Psi(z)$ can be established in a similar 
way to that of the functions $\psi(x)$ [see Eq. $(\ref{e33})$], 
while the vanishing of the integrals $(\ref{e48})$ 
follows---after the interchange of the integral over $t\in [0,1]$ 
and integrals in the $z$--complex plane---from the holomorphy 
of $\log (t  -  z)$ as a function of $z \in \Omega$. The theorem of 
Fubini${}^{11}$ permits this interchange of the integration order
since the function  $F(z,t) \id  \log (t  -  z) 
\varphi(t)$ is in $L^1[\partial \Delta   \times   [0,1]\ ]$
we are interested in. The holomorphy of the functions
 $\Psi^{(j)} (z)$ is an immediate
consequence of the fact that they are  finite sums of 
holomorphic functions.

\vskip 0.3cm 

\centerline{\epsffile [0 0 218 188 ] {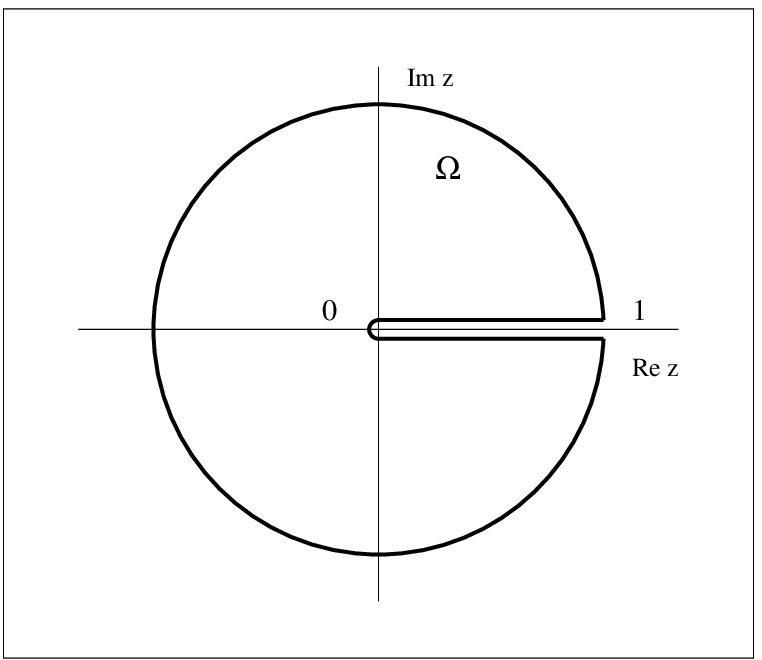}}
\centerline{FIG. 6. {\small The open set $\Omega$.}}

We denote by $\Psi^{\infty}(z)$ the limit of the sequence
$\{ \Psi^{(j)}(z)\}$  wherever this limit exists. We will show
that {\it there are no new
singularities} which enter the region $\Omega$ as a result of the 
summation process, i.e. that the limit $\Psi^{\infty}(z)$ is
holomorphic in $\Omega$. This is an important  point: remind the
counterexample with the geometric series  discussed at the beginning of
the Section $V\ A \ 2$.

In studying the analytic properties of the function
$\Psi^{\infty}(z)$, the crucial property  is again the
uniform convergence---to be proved below---of the sequence
$\{ \Psi^{(j)}(z)\}$.
This might seem surprising since on the 
real line there exist examples of sequences of infinitely 
differentiable functions which converge uniformly to functions
which are nowhere differentiable. However, in the complex plane 
the uniform convergence of the sequences can be used in conjunction
 with the theorem of Morera to prove the holomorphy of their limits.
Indeed, as a first consequence of the uniform convergence 
of the sequence  $\{\Psi^{(j)} (z)\}$ one obtains the continuity
of its limit $\Psi^{\infty}(z)$.  Secondly, the identity
\beq 
\int_{\partial \Delta} \Psi^{\infty}(z) dz = 0
\label{e49}
\eeq 
follows from the vanishing of the corresponding integrals 
over the holomorpic functions $ \Psi^{(j)}(z)$
and from the fact that, because of uniform 
convergence, the integration and the limiting processes
can be interchanged.

To prove the uniform convergence of $\{ \Psi^{(j)}(z) \}$ 
for $z\in \Omega$ we shall
proceed similarly to the method used on the interval $[0,1]$.
Since the sums $(\ref{e47})$ are finite, they  commute with the 
integral $(\ref{e43})$ from the definition of $U_n^{(0)} (z)$ and so
\beq
\Psi^{(j)} (z)\equiv \sum_{n=0}^j
 {\varphi_n \over \lambda_n} U^{(0)}_n(z)=\int_0^1 
\log (t-z) \sum_{n=0}^j \varphi_n u_n(t)dt  .
\label{e50}
\eeq
Using arguments similar to those which led to Eq. $(\ref{e35})$,
we find
\beq
\left| {\textstyle\sum_{j+1}^{j+k} } \psi_n U^{(0)}_n(z) \right|^2
\le \int_0^1 \left| \log(t-z) \right|^2 dt \cdot
{\textstyle\sum_{j+1}^{j+k}}\varphi_n^2 
\label{e51}
\eeq 
where, for all  $z\in \Omega$, the integral over the logarithm 
is bounded while the sum over the coefficients $\varphi_n^2$
tends to zero as $j$ becomes large. This proves the uniform
convergence of the sequence $\{ \Psi^{(j)}(z) \}$ in $\Omega$
and hence, by the theorem of Morera, that  the 
function  $\Psi^{\infty}(z)$ has no singularities 
in the region $\Omega$.

These results can also be proved using a theorem of Vitali${}^{13}$
which states that if:   (a) the functions
$ \Psi^{(j)}(z) $ are holomorphic for $z\in \Omega$,  (b) 
the sequence converges uniformly on some subset $\Sigma$ of $\Omega$
which has an accumulation point inside $\Omega$ 
and  (c) the functions
$ \Psi^{(j)}(z) $ are {\it uniformly bounded} in  $\Omega$, then 
the functions $ \Psi^{(j)}(z) $ tend uniformly towards a function 
$ \Psi^{\infty}(z) $ which is holomorphic in $\Omega$.
Note that the subset  $\Sigma$ may be
the segment just above the real interval $[0,1]$ where
the uniform convergence has been proved in 
Section $ V\ A\ 3$.

Using Vitali's theorem  we can easily find the regions where
$ \Psi^{\infty}(z) $ is holomorphic, by looking at the sets
where the $ \Psi^{(j)}(z) $ are bounded.
In this way we can verify that
unlike the functions $ U{}_n^{(0)} (z) $ which can be
continued on higher Riemann sheets, the limit
$ \Psi^{\infty}(z) $ generally {\it does not exist}
there. This is so because the uniform boundness condition
 (c) is no longer fulfilled outside the first Riemann sheet
(unless the coefficients $\psi_n$ decrease exponentially quickly).
The reason is the existence of the factor 
$\exp(i \lambda_n (z - t))$ in the higher
Riemann sheet continuations $ U{}_n^{(k)} (z) $ of the functions 
$U{}_n^{(0)} (z)$---see Eq. $(\ref{e65})$
from Section  $V\ B\ 4$---which grows exponentially with 
$\lambda_n$ if $(z - t)$ is complex.

Finally we shall show that, similar to  $\psi^{(\infty)}$ on the
real segment, the limit $ \Psi^{\infty}(z) $ coincides inside $\Omega$
with the function $\Psi(z)$ defined in Eq. $(\ref{e46})$. 
This is a direct consequence of the fact that the set
$\hbox{ker \bsf K}$ is empty in the case of  the logarithmic kernel. 
Indeed, replacing $U_n^{(0)}(z)$ in Eq. $(\ref{e47})$ by its 
definition $(\ref{e43})$, Eq. $(\ref{e46})$ and the Schwarz 
inequality give  
\beq
\left | \Psi(z)- \Psi^{(j)}(z) \right |^2 \le  
\left\|\varphi(x)-\sum_0^j \varphi_n u_n(x)  
\right\|^{2}_{L^2} \int_0^1 
\left |\log (t-z) \right |^2 dt 
\label{e52}
\eeq
where the right hand side tends to zero when $j$ grows. $[$For kernels 
other than the logarithmic one with non empty null space, we can 
define  appropriate $\Psi$--functions so that the corresponding 
$\varphi$--functions are contained in $\hbox{ker}^\bot \hbox{\bsf K}$.
The simplest way to do this is to begin with a Neumann
series but stop after few iterations so that the new $\varphi(x)$
should belong itself to $ \hbox {Ran }\hbox{\bsf K}$.$]$ 

\vspace{0.3cm}
\noindent {\bsfl 5. Sums of asymptotic expressions}
 
One of the goals of Section  V B is to derive asymptotic 
expressions valid for $z\rightarrow 0$ for the analytic 
extensions  $U^{(0)}_n(z)$ of the eigenfunctions: 
\beq
 U^{(0)}_n(z)= U^{(0)}_{n,{\mathrm asy}}(z) + U^{(0)}_{n,{\mathrm
rem}}(z) .
\label{e53}
\eeq
The remainder $U^{(0)}_{n,\mathrm rem}(z)$  behaves like
${\cal O}(|z|^{k-\eta})$ where $k$ is some given positive integer
and  $\eta>0$ but otherwise arbitrary small.  However, the 
coefficients of the asymptotic terms $U^{(0)}_{n,\mathrm asy}(z)$ 
contain some (fixed) positive power of the  eigenvalue $\lambda_n$,
depending on the value of the exponent $k$. Since the $\{\lambda_n\}$
tend to infinity with $n$, we should choose carefully
an appropriate definition for the $\{ \Psi^{(j)}_{\mathrm asy}\}$ in
 order to secure their convergence. 

The simplest way to solve this problem is to use the 
iterated integral equations $(\ref{e10})$ discussed in 
Section  III. The eigenfunctions of the iterated kernels 
$K_r(x,t)$ $(\ref{e12})$ are identical with those of the initial one, 
the only change being that the eigenvalues are now 
$(\lambda_n)^r$. This introduces a beneficial
factor $1/\lambda_n^r$ in the coefficients $\psi_{r,n}$ 
of the new functions 
\beq
\psi_r(x) = \int_0^1 K_r(x,t) \varphi(t) dt  ,
\label{e54}
\eeq
\noindent which are now
\beq
\psi_{r,n}={\varphi_n \over \lambda_n^r}  .
\label{e55}
\eeq

\centerline{\epsffile[0 0 288 225] {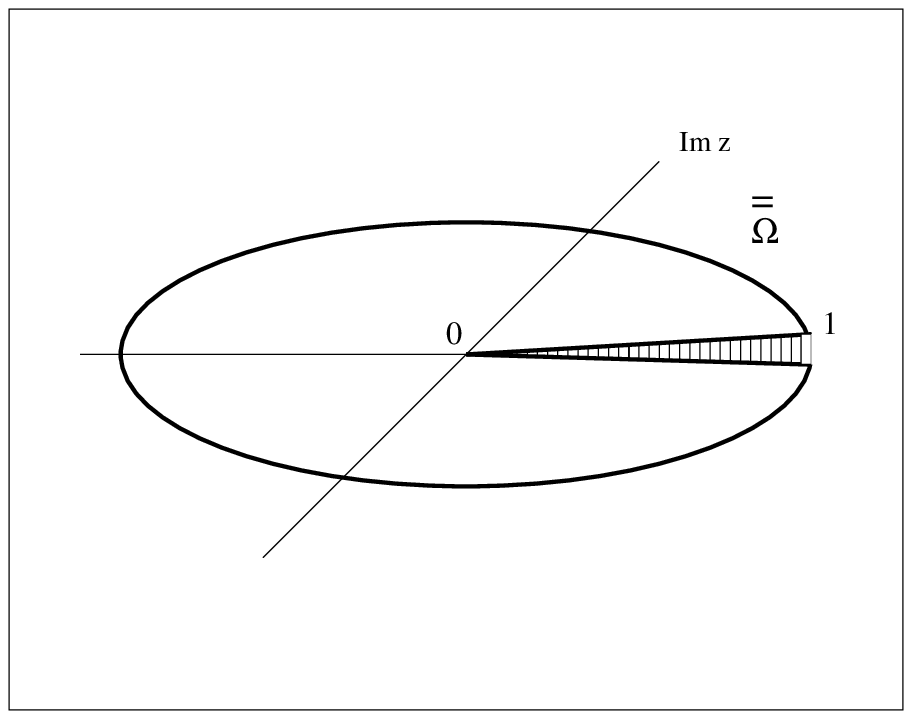}}

\centerline{\vtop {\hsize 11cm {\noindent FIG. 7. {\small 
The closure $\overline {\overline {\Omega}}$ of $\Omega$
drawn in three dimensions in order to emphasize 
that the interval $[0,1]$ is included twice.}}}}
\vskip 0.3cm

\noindent This ensures the separate convergence of the series
\beq
\Psi_{\mathrm asy}^{(j)} (z) = \sum_{n=0}^j \psi_{r,n} U^{(0)}_{n,
{\mathrm asy}}(z)= \sum_{n=0}^j {\varphi_n \over \lambda_n^r} 
U^{(0)}_{n,{\mathrm asy}}(z)
\label{e56}
\eeq
and
\begin{eqnarray*}
\Psi_{\mathrm rem}^{(j)} (z) = \sum_{n=0}^j \psi_{r,n} U^{(0)}_{n,
\mathrm rem}(z)= \sum_{n=0}^j {\varphi_n \over \lambda_n^r} 
U^{(0)}_{n,\mathrm rem}(z)  .
\end{eqnarray*}
As a result the asymptotic expansion of
the solution of the integral equation will contain terms of
the form $z^m \log^k(z)$, $ k\le r$, $ m \ge k$, as do the 
iterated free term [see Eq. $(\ref{e23})$] and the asymptotic 
terms $U^{(0)}_{n,{\mathrm asy}}(z)$.

The initial range of validity of the asymptotic expressions
derived above is the cut open disk $\Omega$ and so does not 
yet extend on the real segment $[0,1]$. We are of course interested 
to show the correctness of these asymptotic series also on some
 small real interval $0\le x \le x_0$.  As has
 been shown above, the theorem of Vitali fails to work 
on higher Riemann sheets since the functions $U^{(1)}_n(z)$ 
and $U^{(-1)}_n(z)$ are no longer uniformly bounded and so
 the sequence $\{ \Psi^{(j)}(z)\}$ does not converge any more
there. However it is interesting and important for what follows
to note that the sequence $\{ \Psi^{(j)}(z)\}$ related to
our integral equation with logarithmic kernel
 converges uniformly also on the closure 
 $\overline {\overline {\Omega}}$ of $\Omega$, i.e. the function
 $\Psi^{\infty}(z)$ is well defined and continuous (because of
the  uniform convergence) on the real interval $[0,1]$, both 
when approached from above and below. 
We have used the symbol $\overline {\overline {\Omega}}$ to 
emphasize the fact that in all this discussion the open set
$\Omega$ has to be considered as an open subset of the  whole
Riemann manifold of the solution of the integral equation 
rather than of the $z$--complex plane (whereas normally the closure 
 $\overline {\Omega}$ coincides with the unit disk $|z|\le   1$).
The set  $\overline {\overline {\Omega}}$ contains the 
interval $[0,1]$ twice (see Fig. 7) corresponding to the fact
that the function $\Psi (z)$ has different limits when $z$ 
approaches the interval $[0,1]$  from above or below.

To prove that the sequence $\{ \Psi^{(j)}(z)\}$ converges uniformly
also on the interval $[0,1]$ it is sufficient to note that when
$z\in [0,1]$,  $\left| \log(t  -   z) \right|$ is equal
 to $\left| \log|t   -   z|\right|$ if $t   >   z$, or to
$\left|\log|t  -   z| \pm i\pi\right|$ if  $t  <z   $. 
Here the sign of $ i\pi$
depends on whether $z$ approaches the real axis from above or
from below. However in both
cases the integral from Eq. $(\ref{e52})$ is bounded and hence,
the right hand side of $(\ref{e52})$ can be made as small as one
wishes by taking $j$ to be sufficiently large. 

It has been already shown in the previous subsection that
limit  $\Psi^{\infty}(z)$ coincides with the holomorphic function 
$\Psi(z)$ throughout the open set $\Omega$. The limit
$\Psi^{\infty}(z)$ does not exist beyond the two real segments
 $[0,1]$ from the border of  $\overline {\overline {\Omega}}$
but is continuous up to and on them, because of the uniform convergence
of the sequence $\{ \Psi^{(j)}(z)\}$. The function $\Psi^{\infty}(z)$
 is hence defined, by continuity, in an unambiguous way 
on the two real intervals on the boundary of
 $\overline {\overline {\Omega}}$. It coincides there with $\Psi(z)$,
as everywhere else in the closed set $\overline {\overline {\Omega}}$.

 In this way we have shown that the asymptotic series 
$\Psi^{\infty}_{\mathrm asy}(z)$ obtained using  the asymptotic 
expansions of $U_n^{(0)}(z)$ are valid in a neighbourhood of the
origin in $\overline {\overline {\Omega}}$, and therefore on the 
(two) real intervals $[0,x_0]$. This means that the remainder
$\Psi^{\infty}_{\mathrm rem} (z)$ of the asymptotic series is 
bounded by ${\cal O} (|z|^{k-\eta})$  on the 
real interval  $[0,x_0]$ as well as in the open set $\Omega$.

\vspace{0.3cm}
\noindent {\bsf B. Continuation of the eigenfunctions to the
complex plane}

In this subsection we shall study the singularities of the 
eigenfunctions. To this end, similar to the analytic extension 
$F^{(0)}$ of the free term from Section  IV we shall introduce
in Section  $V \ B\ 2$ the analytic extensions $U^{(0)}_n$ of
the eigenfunctions $u_n (x)$.

In trying to continue $U^{(0)}_n$ analytically on higher 
Riemann sheets, i.e. in trying to construct the function $U^{(1)}_n$
as we did with the free term in Section  IV, we face a 
specific difficulty  related to the fact that the eigenfunctions 
$u_n (x)$ as they stand,  are defined only on the segment 
$[0,1]$. This means that we could no longer 'hook' the 
integration contour, as we did in Section  IV where the 
weight $w(t)$, being analytic, was well defined not only on the
real segment but also on the various complex 
plane deformations of the initial integration path.

This point will be solved in Section $V \ B\ 3$  where
the eigenfunctions $u_n (x)$ will be expressed as linear combinations 
of $U^{(0)}_n$ and its Riemann sheet continuations.
Another consequence of this fact will be the Volterra 
integral equation which relates $U^{(1)}_n$ to $U^{(0)}_n$
or vice versa (Section $V \ B\ 4$). This Volterra 
integral equation can be solved effectively, providing  
explicit expressions for $U^{(1)}_n$ or $U^{(-1)}_n$
in terms of $U^{(0)}_n$. Finally, in Section  $V \ B\ 5$ 
this integral equation is used in order to derive the 
asymptotic series which describe the singularities of the 
eigenfunctions around the origin.

\begin{figure}[t]
\begin{center}
\input{fig8}
\end{center}
\begin{center} 
{FIG. 8. {\small The eigenfunctions $u_0(x)$ and $u_1(x)$.}}
\end{center}
\end{figure}

\vspace{0.3cm}
\noindent {\bsfl 1. Eigenfunctions of the logarithmic kernel}

\noindent It has been proved in the  Section  $V \ A\ 3$  that, 
in spite of the logarithmic singularity in the integrand,
the eigenfunctions $\{ u_n \}$ defined by
\beq
 u_n (x)= \lambda_n \int_0^1 \log \left|t- x\right|  u_n (t) dt
\label{e57} 
\eeq
are  continuous functions.
The graphs of some of these eigenfunctions  are depicted 
in Fig. 8 and Fig. 9. Although these eigenfunctions are
finite and  look  well behaved at the end points $0$ and $1$  
of the fundamental domain (they are there continuous),
their derivatives are there infinite. It is the aim of the
following subsections to give a full analytic description of the
singularities of $ u_n (x)$ at these end points.

\begin{figure}
\begin{center}
\input{fig9}
\end{center}
\begin{center} 
{FIG. 9. {\small The eigenfunctions $u_2(x)$, $u_3(x)$ and $u_4(x)$.}}
\end{center}
\end{figure}

Notice that initially, in order to compute the eigenfunctions $u_n (x)$
it has been sufficient to take  the range of the variable $x$ to be
the same as  that of the integration variable~$t$. Usually to
compute the function from the left hand side of an integral equation
for values of the variable $x$ outside this initial range,
 we take the so--called Nystr\o m
 continuation, which consists  simply of substituting 
the new values of $x$ in the right hand side of the equation.
However this is  possible {\it only} if the kernel is 
holomorphic in some given region and so
 this method is not applicable in our case because of the modulus 
function inside the logarithm which spoils holomorphy. Thus if in 
Eq. $(\ref{e57})$ we take $x$ to be negative, the left hand side
$"u_n (x)"$ {\it is not} the  analytic continuation of the function 
$u_n (t)$ which appears on the right hand side.

\vspace{0.3cm}
\noindent{\bsfl 2. The functions \mbox{\boldmath $U{}_n^{(0)} (z)$ }
and their analytic continuations 
\mbox{\boldmath$U{}_n^{(\pm k)} (z)$} }

The function $"u_n (x)"$ which was defined in this naive way for
negative values of $x$, may be  used to introduce a new function
$U{}_n^{(0)} (z)$  which is analytic and, although different
from $u_n (x)$ when $x   \in   [0,1]$, 
is intimately related to it. Noticing that for 
$x  <  0$ {\it and}  
$\forall \ t  \in   [0,1] $ we have 
$\log \left| t  -  x \right | =\log (t   -  x) $,
we shall define $U{}_n^{(0)} (z)$ by
\beq
 U{}_n^{(0)} (z) \df \lambda_n \int_0^1 \log (t-z) u_n (t) dt, 
\qqr n=1,2,3,\ldots . 
\label{e58}
\eeq
Taking the cut of $\log(Z)  \equiv   \log (t   -  z)$
along the real negative axis of the complex  
$Z   \equiv   t  -  z$--plane,
$U{}_n^{(0)} (z)$ is a complex function of real type,
$U_n^{(0)} (z)=\overline{U_n^{(0)}} (\bar z) $,
having, as we shall show later, a branch point at the origin 
and a cut lying on the positive real axis. Since we are 
interested in the singularity of the function near 
the origin we shall consider the detailed behaviour of 
$U{}_n^{(0)} (z)$ only in the neighbourhood of 
$z  =   0$. A similar analysis can be 
performed at the point $z  =  1$.

The superscript (${}^{(0)}$ in the present case)
of $U{}_n^{(k)} (z)$ denotes the Riemann sheet under
consideration. Our notation will be such that, by crossing the
cut of $U{}_n^{(k)}$ between $z  =  0$ 
and $z  =  1$ in {\it  an anti--clockwise} way,
the value of the superscript increases by $1$: 
 see, in Fig. 10, the full line $\alpha^{(0)}$--$\beta^{(0)}$
from the sheet $(0)$, which is continued by
the dashed--line  $\alpha^{(1)}$ lying  on sheet~$(1)$.
Conversely, crossing the cut between $0$ and $1$ in the {\it clockwise} 
direction, we pass from the full line $\alpha^{(0)}$ from the
sheet $(0)$ to the dotted--line $\beta^{(-1)}$ from the sheet $(-1)$.

\vskip 0.2cm
\centerline {\epsffile [0 0 236 140] {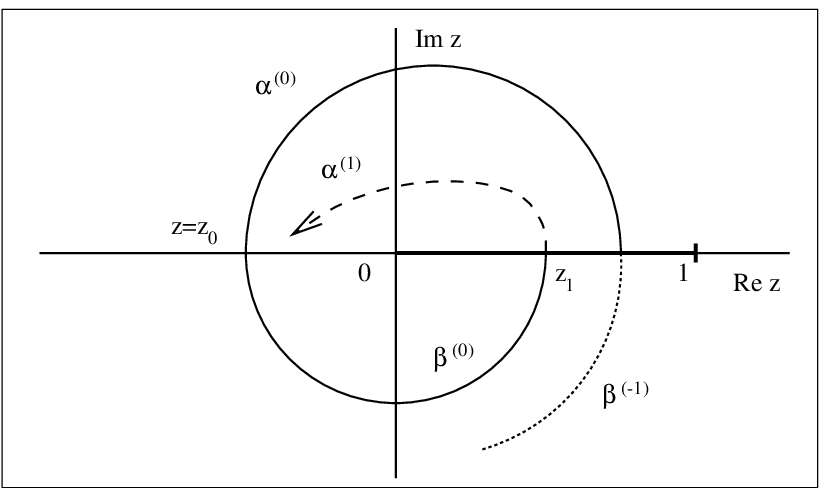}}
\centerline{FIG. 10. {\small Higher Riemann sheet continuations of
$U_n^{(0)}$.}}

We will want to continue the function $U{}_n^{(0)} (z)$ defined 
by the right hand side integral from Eq. $(\ref{e58})$, starting  from 
some point $z  =  z_0$ ($z_0<0$), along the 
full--line path $\beta^{(0)}$ and the dashed--line $\alpha^{(1)}$, 
back to $z_0$, obtaining in this way the value $U{}_n^{(1)} (z_0)$ 
of $U{}_n(z)$ on the next Riemann sheet. But in so doing we will 
need to deform the integration contour on the right hand side of 
Eq. $(\ref{e58})$ into the upper half complex $t-$plane as we did 
in the case of the free term in Section  IV. We are faced, 
however,  with the difficulty that $u_n (t)$, as it stands, is defined 
only on the real segment 
$0  \le  t  \le  1$. 
We therefore need to express  $u_n (t)$ as the value of some analytic 
function on the upper edge of the cut. Here we will be interested 
only in {\it the upper lip} of the cut, since moving in an 
anti--clockwise direction, the analytic continuation path will 
hook and deform  the  integration contour into the {\it upper half} of 
the complex $t$--plane (see the curve {\it C} from Fig. 5).

\vspace{0.3cm}
\noindent {\bsfl 3. Expressing  \mbox{\boldmath $u_n(t)$}, 
\mbox{\boldmath $0<t<1$}, in terms of 
\mbox{\boldmath $U{}_n^{(0)}(t+i\varepsilon )$}  and 
\mbox{\boldmath $U{}_n^{(1)}(t+i\varepsilon )$}} 

 Our first concern will be to rewrite the functions appearing
under the integral sign of Eq. $(\ref{e58})$ as combinations of analytic
functions. To this end we again use  the definition
$(\ref{e58})$ and start from some point $z=z_0$, $z_0 <0$.
We  first continue $U{}_n^{(0)} (z)$ along a path lying
in the lower half $z$--plane to a point 
$z_- =z_1 -i\varepsilon$, $0<z_1 < 1$, $\varepsilon >0$, below 
the segment $[0,1]$ (along the path $\beta^{(0)}$ from Fig. 10). 
If the cut of $\log (Z) \equiv \log (t-z)$ is
taken to run along the negative real $Z$--axis, we have 
$$ \displaystyle
 {\lim_{\varepsilon \down 0 } \log (t-(z_1-i\varepsilon))}=
\cases {\log \left| t-z_1\right| +i \pi & if $t<z_1$, \cr
\log \left| t-z_1 \right| & if $t>z_1$. \cr} $$
We now split the integral along the segment $[0,1]$ into 
one between $0$ and $z_1$ (where, of course, $t<z_1$ and so the
integral runs along the upper lip of the cut of the logarithm), and a 
second one, between $z_1$ and $1$, where the integration points stay 
apart from the cut of the logarithm.  In this way we get
\begin{eqnarray}
 U{}_n^{(0)} (z_1 -i\varepsilon )&=
&\lambda_n \int_0^{z_1}  \log \left| t-z_1 \right|
 u_n (t)dt+\pi i\lambda_n \int_0^{z_1} u_n (t)dt
+\lambda_n \int_{z_1}^1 \log \left| t-z_1 \right| u_n (t)dt \nonumber\\
& =& u_n (z_1)+\pi i\lambda_n \int_0^{z_1} u_n (t)dt 
\label{e59}
\end{eqnarray}
where the last equality follows simply from the fact that the 
first and the third terms  combine to give exactly the
right hand side of Eq. $(\ref{e57})$. 

Similarly we can make an analytic continuation from
$z  =   z_0   <  0$  (along the path $\alpha^{(0)}$ 
from Fig. 10) to the point $z_+   =  z_1  +   i\varepsilon$ 
above the cut, to obtain: 
\begin{eqnarray}
 U{}_n^{(0)} (z_1 +i\varepsilon )&=
&\lambda_n \int_0^{z_1}  \log \left| t-z_1 \right| 
u_n (t)dt-\pi i\lambda_n \int_0^{z_1} u_n (t)dt
+\lambda_n \int_{z_1}^1 \log \left| t-z_1 \right| u_n (t)dt \nonumber\\
& =&u_n (z_1)-\pi i\lambda_n \int_0^{z_1} u_n (t)dt .
\label{e60} 
\end{eqnarray}
The integral over $ u_n (t)$ can be eliminated between 
Eqs. $(\ref{e59})$ and $(\ref{e60})$ by taking the sum of the 
right hand sides, so that
\begin{eqnarray}
 u_n (z_1) 
& =& {1\over 2} [ U{}_n^{(0)} (z_1+i\varepsilon) 
+ U{}_n^{(0)} (z_1-i\varepsilon) ]+ {\cal O} (\varepsilon) 
\nonumber\\ & =& {1\over 2} [ U{}_n^{(0)} (z_1+i\varepsilon) +
 U{}_n^{(1)} (z_1+i\varepsilon) ]+ {\cal O} (\varepsilon) .
\label{e61}
\end{eqnarray}
Here the last line of Eq. $(\ref{e61})$ follows from the fact that the
 values of $ U{}_n^{(1)}$ of the function $ U{}_n$ on the next
 Riemann sheet on the {\it upper lip} of
the cut, merge, by definition (see also Eq. $(\ref{e17})$),
with those of $ U{}_n^{(0)}$ {\it below} the cut: 
$$
U{}_n^{(0)} (z_1-i\varepsilon)= U{}_n^{(1)} (z_1+i\varepsilon) + 
{\cal O} (\varepsilon) \qqr , \qqr 0<z_1<1.$$

\vspace{0.3cm}
\noindent {\bsfl 4. A Volterra equation for 
\mbox{\boldmath $U{}_n^{(1)}(z)$}} 

  We shall now  be able to deform the integration contour in
the complex $t-$plane in the same way as we did in section  IV.
 Recalling that by the analytic continuation of $U{}_n^{(0)} (z)$ 
along the path $\beta^{(0)}$ -- $\alpha^{(1)}$ from Fig. 10 one
obtains the function $U{}_n^{(1)} (z)$, we find
\beq
U{}_n^{(1)} (z_0 )=\lambda_n \int_C \log  (t-z_0 )
 {{U{}_n^{(0)} (t) +U{}_n^{(1)} (t)}\over 2}dt 
\label{e62}
\eeq
where the integration contour ${\it C}$ is shown in Fig.  5 and where 
we have replaced $u_n (t)$ under the integrand by its holomorphic 
expression $(\ref{e61})$. 
The dashed line represents  the cut of $\log  (t-z_0 )$ in  the 
complex $t-$plane, where, again, the {\it $+i\pi$--zone} and 
{\it $-i\pi$--zone} mean that the logarithm differs there by $+i\pi$ and
$-i\pi$ with respect to its mean value across the cut.
In the  region $\hbox {Re }t   < 0$, both $U{}_n^{(0)}$ and 
$U{}_n^{(1)}$ are holomorphic. So---recollect the discussion 
from section  IV which led to Eq. $(\ref{e14})$---the 
integral over the $\hbox {Re }t   <   0$ half-plane part of 
the contour ${\it C}$  yields 
$$-i\pi\lambda_n \int_{z_0}^0 [U{}_n^{(0)}(t)+U{}_n^{(1)}(t)]dt  ,$$
while the rest of the integral, between $t  =  0$ and 
$t  =  1$, is identical with that from definition 
$(\ref{e58})$ of the function $U{}_n^{(0)} (z_0 )$. 
So, Eq. $(\ref{e62})$ can be rewritten in the form
\beq
U{}_n^{(1)} (z )=U{}_n^{(0)} (z) 
+i\pi \lambda_n \int_0^z [U{}_n^{(0)} (t) +U{}_n^{(1)}(t)] dt 
\label{e63}
\eeq
which is valid for any $z$ in the cut complex plane since it 
involves only analytic expressions.  Equation $(\ref{e63})$ can
be regarded as a Volterra integral equation for $U{}_n^{(1)} (z )$
if the function $U{}_n^{(0)} (z )$ is known, or equally, as a Volterra 
integral equation for $U{}_n^{(0)} (z )$ if $U{}_n^{(1)} (z )$
were known.

This equation can be solved by differentiation. We find immediately 
\beq
\frac {dU{}_n^{(1)} (z )}{dz}- \frac {dU{}_n^{(0)} (z)}{dz} =
+i\pi \lambda_n  [U{}_n^{(0)} (z) +U{}_n^{(1)}(z)]  ,
\label{e64}
\eeq
or, rearranging the terms,
\beqa
\frac {dU{}_n^{(1)} (z )}{dz}-i\pi \lambda_n  
U{}_n^{(1)} (z)= \frac {dU{}_n^{(0)} (z)}{dz} 
+i\pi \lambda_n  
U{}_n^{(0)} (z) . \nonumber
\eeqa
This can be rewritten in the form
\beqa
\exp(+i\pi \lambda_n z)
\frac{d}{dz} \Bigl[\exp(-i\pi \lambda_n z) U{}_n^{(1)} (z ) \Bigr]  
=  \exp(-i\pi \lambda_n z)
\frac{d}{dz} \Bigl[ \exp(+i\pi \lambda_n z)U{}_n^{(0)} (z ) \Bigr]  .
\nonumber
\eeqa
Now, from Eq. $(\ref{e63})$ it is obvious  that
$U{}_n^{(1)} (0)  =  U{}_n^{(0)} (0)$. This is an initial condition 
which permits us to write the solution of the Volterra equation as  
\beq
U{}_n^{(1)} (z )=U{}_n^{(0)} (z )+2i\pi \lambda_n \int_0^z 
\exp[i\pi \lambda_n (z-t)] U{}_n^{(0)} (t) dt  .
\label{e65}
\eeq
Similarly we can step backwards and express  $U{}_n^{(0)} $
with respect to $U{}_n^{(1)} $, or 
$U{}_n^{(-1)} $ with respect to  $U{}_n^{(0)}$: 
\beq
U{}_n^{(-1)} (z )=U{}_n^{(0)} (z )-2i\pi \lambda_n \int_0^z 
\exp[-i\pi \lambda_n (z-t)] U{}_n^{(0)} (t) dt  .
\label{e66}
\eeq

\vspace{0.3cm}
\noindent {\bsfl 5. Asymptotic expansion}

To begin with, we shall  suppose  that each of the functions
$U{}_n^{(0)} (z)$ admits  an expansion  around the origin  of the 
form
\beqa
U{}_n^{(0)} (z)&=&U{}_{n,{\mathrm asy}}^{(0)} (z)+U{}_{n, {\mathrm
rem}}^{(0)} (z),
 \nonumber \\
&\equiv &\sum_{k=0}^r a_{k} z^k +\sum_{m=1}^r \sum_{k=1}^r
 b_{mk } z^m \cdot \log^k (-z) +U{}_{n,\mathrm  rem}^{(0)} (z)
\label{e67}
\eeqa
where $U{}_{n,\mathrm rem}^{(0)}$ is ${\cal O} (|z|^{r+1-\eta})$.
The choice of these series may look very restrictive, 
but once we have shown their consistency, their uniqueness follows
from the uniqueness of the solution of a linear  integral equation 
with a Hilbert--Schmidt kernel. In Eq. $(\ref{e67})$ the sum 
$\sum_{k=0}^r a_{k} z^k$ comes from the holomorphic part of
 $U{}_{n}^{(0)}$ 
around $z  =  0$. Taking as before the cut of 
$\log (Z)  \equiv   \log (-z)$ along the negative real $Z$--axis, 
the right hand side of Eq. $(\ref{e67})$ will be holomorphic throughout 
the domain $\Omega$ which has a cut running along the positive real 
axis (see Fig.  6). From  the definition $(\ref{e58})$ we see that  
$U_n^{(0)} (z)$ is a real--analytic function [i.e. $U_n^{(0)} (z)=
\overline{U_n^{(0)}} (\bar z) $] and so  the coefficients 
$a_{k}$ ($k  =  0,1,2 \ldots $) and  
$b_{mk}$ ($m,k  =  1,2,\ldots$) have to be 
\begin{figure}
\begin{center}
\input{fig11}
\end{center}
\begin{center} 
{FIG.11. {\small One parameter asymptotic fit of $\tilde {u}_0(x)$.}}
\end{center}
\end{figure}
real. The coefficients $b_{mk}$ are  then  determined in a 
recursive  manner from the coefficients $a_k$ of the 
regular part. 

In order to find the coefficients $b_{mk}$  we first notice
that the analytic continuation of $U{}_{n,\mathrm asy}^{(0)} $   
across the cut yields
\beq
U{}_{n,\mathrm asy}^{(1)} (z)=\sum_{k=0}^r a_{k} z^k +\sum_{m=1}^r 
\sum_{k=1}^r b_{mk }
 z^m \cdot [ \log (-z) +2i\pi ]^k .
\label{e68}
\eeq
Inserting the expressions $(\ref{e67})$ and $(\ref{e68})$ 
in Eq. $(\ref{e64})$ we may then compare the various terms 
appearing in the left and right hand sides. 
This comparison can be done for any $z$ in the open set 
$\Omega$, for instance for negative real $z$, in order to be 
away from the cut of $\log (-z)$.
Taking the limit $z\rightarrow 0$ in both sides we remark  that  
\beq
b_{1k} = 0, \ k=2,3,\ldots \ ,
\label{e69}
\eeq
as a consequence of the fact that the differences of terms of the type 
$z [\log (-z)  +  2i\pi]^k$ and $z \log^k (-z)$ from the left hand side
 of Eq. $(\ref{e64})$ yield, by differentiation, terms which tend to infinity 
and which are not compensated by similar terms from the right hand side.
Looking at the constant terms we get
\beq
b_{11} =\lambda_n a_{0} .
\label{e70}
\eeq
Similarly, by differentiating with respect to $z$ both sides of 
Eq. $(\ref{e64})$ we obtain
\beq
b_{mk} = 0 \qqr  \hbox{ if} \qqr k\ge m+1   ,
\label{e71}
\eeq
while the coefficients $b_{mk}$ with $k\le m$ can be expressed 
iteratively by means of the coefficients $a_k$. So,
for instance, we find 
\beq
b_{21}=\frac{2a_1 \lambda_n  -a_0 \lambda_n^2}{4}  , 
\qqr  b_{22}=\frac{a_0 \lambda_n^2}{4} .
\label{e72}
\eeq
We tried to check numerically the accuracy of this asymptotic
expansion in regions close to the singularity.
To this end we took two terms from the
holomorphic part of $U{}_{n,\mathrm asy}^{(0)} (z)$, and expressed the
corresponding $b_{mk}$ coefficients in terms of
the first two coefficients $a_0$ and $a_1$. Since $a_0$ is determined
by the value at $x  =   0$ of the eigenfunctions which, in turn, 
is determined by the normalization condition, we have in fact {\it only
one free parameter}, the coefficient $a_1$. Using the above
expressions for $b_{11}$, $b_{21}$ and $b_{22}$ in terms of $a_0$
(given) and $a_1$ (free), we have chosen $a_1$ to obtain the best
fit to the first few eigenfunctions, in the region $[0,0.01]$
near the origin where we expected the asymptotic expansion to hold.
\begin{figure}
\begin{center}
\input{fig12}
\end{center}
\begin{center} 
{FIG.12. {\small One parameter asymptotic fit of $\tilde {u}_1(x)$.}}
\end{center}
\end{figure}
When we plot the function $u_n (x)$ together with our asymptotic 
expansion the two curves appear identical on the interval $[0,0.01]$. 
In order to show the slight difference in these functions we have
defined a new function
${\tilde u}_n (x)  =  u_n(x)  -  u_n(0)  -  
(u_n(x_1)  -  u_n(0))x/x_1$ 
which has its end points ${\tilde u}_n (0)$ and ${\tilde u}_n (0.01)$
at the same height and so permits the use of a much enlarged $y$--axis
scale. The corresponding plots for the two first eigenfunctions
are presented in Figs. 11 and 12.
Although as it has been already stressed,
these curves are just one  parameter fits and that moreover
we have restricted ourselves only to the first terms in the 
asymptotic expression, the agreement between these asymptotic 
expressions (the full lines) with the computer
calculated points of the eigenfunctions (the crosses) 
is really excellent.

\vspace{0.3cm}
\noindent {\bsf VI. \qqr RESUME AND CONCLUSIONS}

As indicated in the Introduction, the numerical calculations
related to the solution of the inverse problem 
for EIT are seriously hampered by the 
high number of mesh points necessary to take into account the sharp
peaks of the current density near the edge of the electrodes. Since 
these peaks seemed to be intrinsic objects  describable by only a  small
numbers of parameters we have investigated the details of their 
analytic structure. 
To this end we have studied the Riemann sheet structure of 
the eigenfunctions of the dominant singular  integral equation relating 
to the solution of the mixed boundary problem for the potential, and
derived asymptotic expressions both for the eigenfunctions and for the
solution of the integral equation. These asymptotic 
expressions provide very
simple parametrisations for the anomalous thresholds, whose
effectiveness can be judged from the Figs. 11 and 12.

The paper is constructed as follows:  In the Introduction and in 
Section  II the mathematics of the EIT modelling is discussed 
while the corresponding weakly singular integral equation is
given in  Section  III. In Section  IV we have discussed
the effect on the singularities of 
the free term of the moving cut of the logarithm which
'hooks' the integration contour. The {\it generic singularities}
of the solution of the integral equation are then
described as a superposition of those of the free term and
of the eigenfunctions which were derived in Section  V B.
Section  V A contains
the proof that although we deal with infinite series,
no new singularities appear in the neighbourhood of the origin,
i.e. the singularities of the solution are really those of the 
eigenfunctions and of the free term. We also determine 
the limits of the domain where the eigenfunction sum 
converges unconditionally and show how one can extend 
the validity of the asymptotic series also on the real segment 
$[0,1]$.

We hope that the discussion of the analytic properties of the
eigenfunctions and solution of this quite special, logarithmic singular
equation, will  provide a working example which might be useful also 
for the study of the singularities of the solution of other weakly 
singular integral equations. The discontinuities across the cuts 
will certainly be different, but the general discussion will 
probably be fairly similar.

\vskip 0.8cm

\noindent {\bsf   ACKNOWLEDGEMENTS}

The authors are grateful to Professors G. Auberson, G. Mennessier
and P. C. Sabatier for long discussions concerning the 
integral equation and other analytic aspects of this paper. One of
authors (M.K.P.) would like to thank The British Council for its 
support. Another author (S.I.) acknowledges a research grant from 
the Minist\`ere  de l'Enseignement Sup\'erieur et de la Recherche 
and the help of Professor
C. Duhamel from the French Embassy in Bucharest without whom this
collaboration would not have been possible.

\newpage
\vspace{0.3cm} 
\noindent {\bsf APPENDIX:  COMPLETENESS OF THE BASIS 
\mbox{\boldmath $\{ u_n \}$}}

Following a proof given by G. Auberson,${}^{14}$ we shall show 
 in what follows  that $\hbox{ker \bsf K}$, the null space of the 
logarithmic kernel, is empty and so the  eigenfunctions
 $\{ u_n \} _{n=0,1,2 \ldots} $ do span the whole Hilbert space
of the $L^2$ functions on $[0,1]$.

Suppose that $\hbox{ker \bsf K}\not = \{ 0 \}$, i.e. that there
exists at least one non zero $L^2$--function $\phi$ such that
$$
\int_0^1 \log \left| x-y \right| \phi (y) dy =0  .\eqno (\hbox{A1})
$$
Defining the function
$$
v (x)\df \int_0^x   \phi (y) dy  , \eqno (\hbox{A2})
$$
we have
\beqa
&a)&  \frac{d v(x)}{dx}=\phi (x) \qqr \hbox { almost everywhere}, 
\nonumber \\
&b)&  v(0)= 0 . \nonumber
\eeqa
We can easily prove that the function $v(x)$  
satisfies a H\"older condition of 
index $1/2$
$$
 \left| v(x_1) - v(x_2)\right| \le A \left| x_1 - x_2 \right|^{1/2} \ ,
\hbox { for } \forall x_1,x_2 \in
(0,1)\ ,$$
where $A$ is a positive constant.

By integrating the left hand side of  Eq. (A1) by parts we find 
$$\int_0^1 \log \left| x-y \right| \phi (y) dy =v(1) \log (1-x)- 
{\cal P}\int_0^1 \frac{v(y)}{y-x} dy \  $$
so that from Eq. (A1) we obtain 
$${\cal P}\int_0^1 \frac{v(y)}{y-x} dy = v(1) \log (1-x)  ,
\eqno (\hbox{A3})$$

If we now consider the following function
$$ F (z) \df \sqrt{z(z-1)} \int_0^1 \frac{v(y)}{y-z} dy \qqr \hbox 
{  for } 
\qqr z\in D \eqno (\hbox{A4})$$
where $D$ is the complex $z$--plane cut along the  
segment  $[0,1]$,
we can show that: 
\beqa
&(i)&  F \ \hbox { is a holomorphic function in } D  ;\nonumber \\
&(ii)&  \lim_{z\rightarrow \infty}  F(z) = -  \int_0^1 v(y)  dy ;
 \nonumber \\
&(iii)& \hbox {Im } F(x+i\varepsilon)=\sqrt{x(1-x)} 
{\cal P} \int_0^1 \frac{v(y)}{y-x} dy
=\sqrt{x(1-x)} v(1) \log (1-x), \hbox { for } 
 x\in (0,1) ;\nonumber\\
&(iv)& \hbox {Re }F(x+i\varepsilon)=-\pi \sqrt{x(1-x)} v(x)  , 
 \hbox { for }  x\in (0,1)  .\nonumber
\eeqa
The properties $(i)-(iii)$ imply that
$$ F(z)= \frac {v(1)}{\pi}  \int_0^1 \frac {\sqrt {y(1-y)}
 \log (1-y)}{y-z}   dy
 - \int_0^1  v(y) dy  , \qqr z\in D  , \eqno (\hbox{A5})$$
while from  $(iv)$ it follows that for $x\in (0,1)$ we have 
$$ -\pi \sqrt{x(1-x)} v(x) = \frac {v(1)}{\pi}  {\cal P} \int_0^1 
\frac{\sqrt{y(1-y)}\log (1-y)}{y-x} dy - \int_0^1  v(y) dy   . 
\eqno (\hbox{A6})$$
Taking now the limits $x{\scriptstyle\searrow} 0$ and
 $x {\scriptstyle\nearrow} 1$ we find
$$ 
\qqr \  \frac {v(1)}{\pi} \int_0^1  \sqrt  
{\frac {1-y}{y}} \log (1-y)  dy - \int_0^1 v(y) dy  = 0  , 
\eqno (\hbox{A7})$$
$$-\frac {v(1)}{\pi} \int_0^1   \sqrt 
{\frac {y}{1-y}} \log (1-y) dy - \int_0^1 v(y) dy  = 0  . 
 \eqno (\hbox{A8})$$ 
Subtracting (A8) from (A7) we see that $v(1) = 0$ and 
hence, from Eq. (A6), 
$$v(x)= \frac{1}{\pi \sqrt{x(1-x)}} \int_0^1  v(y) dy \equiv \frac{C}
{\pi \sqrt{x(1-x)}} .\eqno (\hbox{A9})$$
Now, since $v(1)$ is zero, the constant $C$ has to be zero too and 
so $v(x)$ has to vanish identically [it had to be so since neither
the right hand side of Eq. (A9) and even less its derivative 
are $L^2$]. This  implies that $\hbox{ker \bsf K}$ is an  empty 
set and so the eigenfunctions $\{ u_n \}$ of the logarithmic 
kernel form a complete $L^2$ basis on the segment $[0,1]$.

\vskip 0.5cm

{\small \noindent ${}^1$ See for example the collection
of papers in Physiol. Meas., {\bf 16}, Supplement {\bf 3A}, (1995).

\noindent ${}^2$ K.S. Paulson, W.R. Breckon and  M.K. Pidcock,
'Electrode modelling in Electrical Impedance Tomography', SIAM J.
Appl. Math., {\bf 52}, pp. 1012--1022, (1992).

\noindent ${}^3$ E. Sommersalo, M.  Cheney and D. Isaacson, 
'Existence and uniqueness for electrode models for Electric 
Current Computed Tomography', SIAM J. Appl. Math.,
{\bf 52}, pp. 1023--1041, (1992).

\noindent ${}^4$ M.K. Pidcock, S. Ciulli and S. Ispas,
'Singularities of mixed
boundary value problems in Electrical Impedance Tomography',
Physiol. Meas., {\bf 16}, pp. 213--218, (1995).

\noindent ${}^5$ K. Cheng, D. Isaacson, J.C. Newell and D.G. Gisser,
'Electrode models for electric current computed tomography',
IEEE Trans. Biomed. Engrg., {\bf 36}, pp. 918--924, (1989).

\noindent ${}^6$ S. Ciulli, S. Ispas and  M.K. Pidcock, 'Numerical
modelling of a mixed  Neumann--Robin boundary value problem', PM--95/22,
submitted for publication (1996).

\noindent ${}^7$ R.J. Eden, P.V. Landshoff,   D.I. Olive 
and J.C. Polkinghorne, {\it The analytic S--matrix}, (Cambridge
University, Cambridge, 1966).

\noindent ${}^8$ see for instance M. Ciulli,   S. Ciulli  and 
T.D. Spearman, 'Bounds for the Continuation of Perturbative Results 
in the Spectral Region', J. Math. Phys. {\bf 25}, pp. 3194--3203, 
(1984).

\noindent ${}^9$ F.D. Gakhov, {\it Boundary Value Problems}, 
(Pergamon, London, 1966).

\noindent ${}^{10}$ S. Ciulli, S. Ispas and  M.K. Pidcock ---
'Riemann sheet structure of eigenfunctions for a boundary value 
problem related to Electrical Impedance Tomography', 
in preparation (1996).

\noindent ${}^{11}$ W. Rudin,  {\it Real and Complex Analysis},
(McGraw--Hill, New York, 1970).

\noindent ${}^{12}$ J. W. Dettmann,  {\it Applied Complex Variables},
(MacMillan, New York, 1965).

\noindent ${}^{13}$  E. C. Titchmarsh, {\it The Theory of Functions}
2nd ed. (Oxford University Press, London, 1939).

\noindent ${}^{14}$ G. Auberson,  private communication.

}


 \end{document}